\documentclass[twocolumn,superscriptaddress,floatfix,preprintnumbers,prd ,nofootinbib,hyperref]{revtex4-2}
\usepackage[colorlinks=true,breaklinks=true]{hyperref}
\usepackage{comment}
\usepackage[utf8]{inputenc}
\hypersetup{allcolors=[rgb]{0.0 0.0 0.6},linkcolor=[rgb]{0.75 0.05 0.05}}
\usepackage{mathtools}
\usepackage[dvipsnames]{xcolor}
\usepackage{float}
\usepackage{letltxmacro}
\LetLtxMacro{\oldcite}{\cite}
\renewcommand{\cite}[1]{\mbox{\oldcite{#1}}}

\usepackage{setspace}

\usepackage{orcidlink}
\usepackage{enumitem}

\long\def\exclude#1{}

\newcommand{\GE}{\Gamma_{\rm E}}
\newcommand{\GA}{\Gamma_{\rm A}}
\newcommand{\vs}{v_{\rm s}}
\newcommand{\xs}{\xi_{\rm s}}
\newcommand{\rs}{r_{\rm s}}
\newcommand{\zs}{z_{\rm s}}

\bibpunct{[}{]}{,}{n}{}{,}

\begin{document}

\title{Supernova Emission of Secretly Interacting Neutrino Fluid: Theoretical Foundations}

\author{Damiano F.\ G.\ Fiorillo \orcidlink{0000-0003-4927-9850}} 
\affiliation{Niels Bohr International Academy, Niels Bohr Institute,
University of Copenhagen, 2100 Copenhagen, Denmark}

\author{Georg G.\ Raffelt
\orcidlink{0000-0002-0199-9560}}
\affiliation{Max-Planck-Institut f\"ur Physik (Werner-Heisenberg-Institut), F\"ohringer Ring 6, 80805 M\"unchen, Germany}

\author{Edoardo Vitagliano
\orcidlink{0000-0001-7847-1281}}
\affiliation{Racah Institute of Physics, Hebrew University of Jerusalem, Jerusalem 91904, Israel}

\date{July 31, 2023}

\begin{abstract}
Neutrino-neutrino scattering could have a large secret component that would turn neutrinos within a supernova (SN) core into a self-coupled fluid. Neutrino transport within the SN core, emission from its surface, expansion into space, and the flux spectrum and time structure at Earth might all be affected. We examine these questions from first principles. First, diffusive transport differs only by a modified spectral average of the interaction rate. We next study the fluid energy transfer between a hot and a cold blackbody surface in plane-parallel and spherical geometry. The key element is the decoupling process within the radiating bodies, which themselves are taken to be isothermal. For a zero-temperature cold plate, mimicking radiation into free space by the hot plate, the energy flux is 3--4\% smaller than the usual Stefan-Boltzmann Law. The fluid energy density just outside the hot plate is numerically 0.70 of the standard case, the outflow velocity is the speed of sound $\vs=c/\sqrt{3}$, conspiring to a nearly unchanged energy flux. Our results provide the crucial boundary condition for the expansion of the self-interacting fluid into space, assuming  an isothermal neutrino sphere. We also derive a dynamical solution, assuming the emission suddenly begins at some instant. A neutrino front expands in space with luminal speed, whereas the outflow velocity at the radiating surface asymptotically approaches $\vs$ from above. Asymptotically, one thus recovers the steady-state emission found in the two-plate model. A sudden end to neutrino emission leads to a fireball with constant thickness equal to the duration of neutrino emission. 
\end{abstract}

\maketitle

\tableofcontents
\section{Introduction}

Neutrino-neutrino scattering is notoriously difficult to explore experimentally, leaving room for the speculation of a potentially large modification by a new force carrier and thus to large neutrino secret interactions 
($\nu$SI)~\cite{Berryman:2022hds}. Motivated by the neutrino observation of the historical SN~1987A, the idea has taken root that supernova (SN) physics and the detectable neutrino signal should be good places to search for such effects \cite{Dicus:1982dk, Gelmini:1982rr, Kolb:1987qy, Manohar:1987ec, Berezhiani:1987gf, Dicus:1988jh, Fuller:1988ega, Berkov:1988sd,Konoplich:1988mj, Berezhiani:1989za, Farzan:2002wx, Blennow:2008er, Heurtier:2016otg, Das:2017iuj, Shalgar:2019rqe, Chang:2022aas, Fiorillo:2022cdq,Akita:2022etk, Cerdeno:2023kqo,Manzari:2023gkt}. If neutrinos in a SN core form a self-coupled fluid instead of an ideal gas, for sure one would expect dramatic modifications of the diffusive transport out of the SN core, the expansion into space, and the detectable signal properties.

A systematic treatment of these questions must include several elements. One is the radiative transfer of energy and flavor-dependent lepton number by neutrinos, both in the diffusion limit deep inside and in the decoupling region, the neutrino sphere, where the neutrino optical depth in the nuclear medium becomes small. At larger distances, one has to deal with the free expansion of a self-interacting fluid into space, a purely hydrodynamical problem, where the relativistic velocity of sound $\vs=c/\sqrt{3}\simeq0.577\,c$ is a crucial scale. (Henceforth we will use natural units with $\hbar=c=k_{\rm B}=1$.)

Some previous works have focused on this hydrodynamical question alone, modeling the SN neutrino burst as a self-interacting fluid that expands freely after it has been released.  Many years ago, Ref.~\cite{Dicus:1988jh} has posed the problem as that of a gas in plane-parallel geometry, which is suddenly left free to expand after a piston is removed. Very recently, Ref.~\cite{Chang:2022aas} has returned to this subject and studied the same question in spherical geometry. They proposed a self-similar solution for the freely expanding fluid with no interaction with matter, akin to an impulsive energy release. 

A SN burst with a duration of, say, 3~s has a length of $10^6$~km, compared with the proto neutron star (PNS) radius of 10~km. Obviously, the sudden release of the fluid does not correspond to the physics of realistic neutrino emission. Therefore, while the burst solutions of Refs.~\cite{Dicus:1988jh, Chang:2022aas} are interesting pedagogical problems to study the behavior of a freely expanding blob of relativistic fluid, similar to earlier studies in a different context \cite{vitello1976hydrodynamic, yokosawa1980relativistic}, they do not include the physics of neutrino emission by a quasi-thermal steady source that feeds neutrino emission over a period very long compared with all other relevant scales. Therefore, we completely dismiss such solutions in the context of PNS cooling.

The authors of Ref.~\cite{Chang:2022aas} have proposed a second scenario, corresponding to steady outflow. Specifically in their Appendix~H, they have worked out the hydrodynamic solution outside of the PNS and have matched it to a stationary solution within the PNS. At the same time, they questioned whether such a solution could actually be realized, or whether it required special conditions. The reasons for such doubts were left unclear. It is clear that the PNS acts as a source and thus as a boundary for the neutrino outflow. As we will see, we recover the picture of a steady wind, lasting for several seconds, and propagating into space.

As a first approximation, one can picture neutrinos being emitted as blackbody radiation from the PNS surface. The burst duration (or spatial profile) is set by the speed of PNS cooling, which is essentially set by convective transport, and the decoupling process at the surface. But irrespective of the detailed physics of energy transfer, the PNS acts as a huge thermal reservoir of energy to be emitted in neutrinos. Typically, a SN emits 200--400~B (1~B $=$ 1~bethe${}=10^{51}$~erg) in neutrinos. On the other hand, assuming an internal temperature of 30~MeV and a radius of 10~km, the energy stored in a gas of six species of nondegenerate neutrinos in the PNS volume is around 1.2~B. Therefore, the physics of PNS cooling requires the steady production of neutrinos to be emitted in the decoupling region, not simply the diffusion of trapped neutrinos out of the PNS. Even if neutrino diffusion transport (and not convection) dominates heat transfer within the PNS, for our purposes, the latter acts as a traditional blackbody source of radiation, without having to specify the detailed internal physics of the emitter.

Therefore, here we treat the PNS as a source that generates neutrinos and acts as a heat reservoir for their thermal emission. After an initial transient emission from the PNS surface, the solution near the PNS quickly relaxes to a steady state analogous to the steady wind of Ref.~\cite{Chang:2022aas}, although it differs inside the PNS, where the nucleons feed energy to the neutrino fluid.
Actually, starting the emission process at some initial time, the escaping neutrino-fluid front moves with the speed of light, and thus supersonically, so that the fluid near the PNS cannot know what happens far away. The emission from the PNS surface can be understood locally and does not require the dynamical solution far away from the PNS.

One can mimic this situation with the neutrino-fluid energy transfer between two blackbody surfaces at different temperatures that could have plane-parallel geometry or two nested shells in spherical geometry. In the limit of vanishing temperature for the cold surface, the steady-state solution mimics thermal emission into free space without the dynamical approach to an asymptotic steady-state solution. In any case, we find that the dynamical expansion with a sudden initial beginning of neutrino emission approaches asymptotically the steady-state solution near the surface.

If we picture the PNS as a thermal source, the main challenge is to understand the steady release of the neutrino fluid from the PNS surface. If the energy flux from deep inside to the surface is carried by neutrinos, the first question is how diffusive radiative transport in the hot nuclear background medium is modified by large $\nu$SI. If we consider a background medium with a prescribed temperature gradient in plane-parallel geometry, we find that the diffusive transport of energy is exactly the same for standard radiation or a fluid, with the caveat that the neutrino average mean-free path (MFP) against absorption on nucleons differs from the usual Rosseland mean. To reach this conclusion, we assume that the MFP $\lambda_{\nu N}$ does not depend on energy; we outline in Appendix~\ref{app:energy_dependent} how the approach should change to account for the energy dependence of the MFP. 

The next question, and the main subject of our paper, is the decoupling of the radiation from the surface, i.e. the transition between diffusion and free expansion. To study this regime, we use the model of an isothermal body (temperature $T$) that ends abruptly at its surface. For standard (i.e.\ without secret interactions) radiation, the emerging energy flux is given by the Stefan-Boltzmann Law that the energy flux is $F=e_\mathrm{eq}/4$, where $e_\mathrm{eq}$ is the blackbody energy density prescribed by the radiator's $T$. One factor 1/2 comes from the fact that only the outward-going modes are occupied outside of the radiating surface and a factor 1/2 from the angle average of the speed of different modes away from the emitting surface.

In the fluid case, i.e.\ with secret interactions among neutrinos, energy and momentum conservation is enough to find the solution that interpolates between thermal equilibrium deeply inside and free expansion outside, a transition taking place over a few $\lambda_{\nu N}$. Surprisingly, the usual thermal flux is reduced by only 3--4\%, which
arises from the near-cancellation of two competing effects.
In the standard case, the energy density outside the radiating surface is $0.50\,e_\mathrm{eq}$ because only the outward-moving modes are occupied. In the fluid case, this is found to be roughly $0.35\,e_\mathrm{eq}$. In the standard case, the average speed of the modes away from the surface is $0.50$. In the fluid case, the outflow velocity is $\vs=1/\sqrt{3}$, which implies an energy flux $4\vs/(3+\vs^2)=2\sqrt{3}/5\simeq0.69$ times the laboratory-frame energy density $0.35\,e_\mathrm{eq}$ and thus $F\simeq0.24\,e_\mathrm{eq}$. We have not found a fundamental reason why the thermal flux of a neutrino fluid vs.\ standard neutrino radiation should be so similar.

Looking at the problem from several perspectives, we always find that standard radiation and a self-coupled fluid yield surprisingly similar results for radiative energy transport and the emission from a hot body.

To substantiate these findings, we begin in Sec.~\ref{sec:Diffusion} with the question of radiative transfer in the diffusion limit by a self-coupled fluid instead of radiation. Next we turn, in Sec.~\ref{sec:plates}, to the steady-state energy transfer between two blackbody plates of different temperature, considering both plane-parallel and spherical geometry. For comparison, in Sec.~\ref{sec:dynamical} we also derive the dynamical fluid expansion if the blackbody emission starts suddenly at some instant. In Sec.~\ref{sec:End} we study the opposite problem of the signal suddenly switching off and recover the usual conclusion that, in spherical geometry, most of the neutrino burst propagates as a shell of constant thickness. In Sec.~\ref{sec:discussion} we conclude with a discussion and outlook as to what our findings might mean for more realistic SN physics with large $\nu$SI.

\section{Diffusive Energy Transfer}

\label{sec:Diffusion}

Before neutrinos can be emitted from the PNS surface, the energy must be dredged up from deeper inside. While one often reads as an explanation for the PNS cooling time scale that the main agent is neutrino radiative transfer, Ledoux convection plays a major role \cite{Epstein1979, Burrows+1988, Keil+1996, Janka+2001proc, Dessart+2006, Nagakura+2020} and defines the cooling time scale of a few seconds. However, we here suppose that neutrino radiative transfer is important and ask for the modifications by large $\nu$SI. Of course, the efficiency of radiative transfer also determines if convection occurs in the first place, and so the possible modifications need to be understood.

We here consider a region so deep inside the PNS that radiative transport is diffusive, meaning that the effective MFP is short compared with the spatial scale of the radiation-density variation: $|d\log e_\mathrm{eq}/dr|^{-1}\gg \bar\lambda$, where $e_\mathrm{eq}(r)$ is the blackbody radiation density engendered
by local thermal equilibrium (LTE), i.e., $e_\mathrm{eq}(r)$ is a property of the medium. The neutrino energy flux at radius $r$ is in the diffusion limit~\cite{Mihalas:1978, Rutten:2003, chandrasekhar2013radiative}
\begin{equation}\label{eq:Diffusive-Flux}
    F(r)=-\frac{\bar\lambda(r)}{3}\,\nabla e_\mathrm{eq}(r).
\end{equation}
This result follows from the radiation being slightly disturbed from LTE, sporting a small anisotropy that leads to an energy flux that depends linearly on the MFP. The question is if and how $\bar\lambda$ changes by large $\nu$SI. 

\subsection{Standard neutrinos}

The main neutrino interaction channels in a PNS are with nucleons. There are charged-current (CC) processes of the type $\nu_e n\leftrightarrow p e$, $\bar\nu_ep\leftrightarrow ne^+$, and analogous for the $\mu$ flavor, although the latter have been included only in some recent numerical models \cite{Bollig:2017lki, Fiorillo:2023-SN1987A}. The large muon mass of $105.66$~MeV suppresses them in the colder SN regions. Neutral-current (NC) processes of the type $\nu N\leftrightarrow\nu N$ apply to all flavors, although the cross section is smaller than for CC processes, so the latter dominate whenever they are available. Moreover, NC collisions, to lowest order, preserve energy and have the main effect of changing the $\nu$ direction, not its energy. For the $\nu_\tau$ flavor and for $\nu_\mu$ in regions with few muons, (inverse) bremsstrahlung $NN\leftrightarrow NN\nu\bar\nu$ in the main mode for energy exchange. Energy exchange is also achieved in purely leptonic processes, or processes involving pions, but nuclear bremsstrahlung is the main effect \cite{Janka:2012wk, Janka2017Handbooka, Burrows+2020, Mezzacappa+2020, Burrows+2021}.

CC processes as well as bremsstrahlung are absorptive and as such apply separately to neutrinos of any energy~$\epsilon$. For the occupation number $f$ (assumed at energy $\epsilon$) of the neutrino radiation field, the Boltzmann Collision Equation (BCE) for propagation along the $z$-direction is
\begin{eqnarray}\label{eq:BCE-stationary}
\frac{\partial f}{\partial t}+\cos\theta\,\frac{\partial f}{\partial z}&=&-\GA f+\GE(1-f)
\nonumber\\
&=&-\underbrace{(\GA+\GE)}_{\textstyle\Gamma}f+\GE,
\end{eqnarray}
where the propagation angle $\theta$ is relative to the $z$ direction (homogeneity in the other directions assumed), $\GA$ is the absorption rate by the background medium, and $\GE$ the spontaneous emission rate that is reduced by a Pauli-blocking factor. For bosons, $(1+f)$ would appear instead. In the second line, $\GE$ was included as a negative absorption rate in $\Gamma$, which is the traditional ``reduced'' absorption rate, although it is actually enhanced. The terminology comes from photon (boson) radiative transfer, where the sign is opposite \cite{Caputo:2022rca}.

In a stationary situation and when the radiation is in LTE with the surrounding thermal nuclear medium and in the absence of gradients, the neutrinos will follow a Fermi-Dirac distribution of the form
\begin{equation}
    f=\frac{1}{e^{(\epsilon-\mu)/T}+1}
\end{equation}
with the neutrino energy $\epsilon$, chemical potential $\mu$, and temperature $T$. In LTE, the left-hand side of Eq.~\eqref{eq:BCE-stationary} vanishes and the two coefficients are related by
\begin{equation}
    \GA=e^{(\epsilon-\mu)/T}\GE,
\end{equation}
the detailed-balance relation between emission and absorption by a thermal medium. Notice that the medium not only has a temperature $T$, but also a would-be chemical potential $\mu$ for the neutrinos which are not part of the medium. Finally we obtain
\begin{equation*}
    \Gamma=\left[1+e^{-(\epsilon-\mu)/T}\right]\GA
\end{equation*}
for the reduced absorption rate which is {\em the} absorption rate appearing the BCE.

The neutrino gas will be near LTE so we can expand the distribution function in the form $f=f^{\rm th}+\delta f$ so that most terms of the thermal distribution drop out. However, we have to be careful about the perturbative expansion to keep terms of the same order. As per our diffusion assumption, the absorption rate $\Gamma$ is large compared with the gradient of the radiation field. Therefore, expanding consistently to first order in small quantities yields the stationary BCE
\begin{equation}
    \cos\theta\,\frac{\partial f^{\rm th}}{\partial z}=-\Gamma\,\delta f,
\end{equation}
where $\Gamma$ is the monochromatic reduced absorption rate for neutrino energy $\epsilon$. We can directly invert this equation and find for the disturbance
\begin{equation}
    \delta f=-\lambda_{\nu N}\cos\theta\,\frac{\partial f^{\rm th}}{\partial z},
\end{equation}
where here and henceforth we use $\lambda_{\nu N}=1/\Gamma$ for the MFP corresponding to the reduced absorption rate and thus to the reduced opacity. So the lowest-order deviation from a thermal distribution is a small dipole in the angular distribution.

We can now write the monochromatic energy flux, differential with regard to neutrino energy $\epsilon$,
\begin{eqnarray}\label{eq:monochromatic}
\frac{dF(\epsilon)}{d\epsilon}&=& -\lambda_{\nu N}(\epsilon)
\int_{-1}^{+1} \frac{2\pi\,\epsilon^2\,d\cos\theta}{(2\pi)^3} \epsilon \cos^2 \theta\, \frac{\partial f^\mathrm{th}_\nu}{\partial z }
\nonumber\\
&=&-\frac{\lambda_{\nu N}(\epsilon)}{3}\,\frac{\partial e_{\mathrm{eq},\epsilon}}{\partial z},
\end{eqnarray}
where $e_{\mathrm{eq},\epsilon}$ is the blackbody spectral intensity normalized such that $\int d\epsilon e_{\mathrm{eq},\epsilon}=e_\mathrm{eq}$ is the blackbody radiation density in LTE at the local conditions. 

If the MFP does not depend on energy, called the ``gray approximation'' in radiative transfer, one immediately arrives at the form of Eq.~\eqref{eq:Diffusive-Flux}.

However, neutrino cross sections strongly vary with energy. If we
were to suppose that in the relevant SN region their distribution is not strongly degenerate and that the energy flux is essentially driven by the temperature gradient, one can arrive at the integral form of Eq.~\eqref{eq:Diffusive-Flux} such that $\bar\lambda$ has the standard meaning of a Rosseland average. To this end we write the rhs of the monochromatic form Eq.~\eqref{eq:monochromatic} as $-(\lambda_{\nu N}/3)(\partial e_{\mathrm{eq},\epsilon}/\partial T)\nabla T$. After integrating over energy, this yields the integrated flux
\begin{equation}
    F=-\frac{\nabla T}{3}\,
    \int d\epsilon\,\lambda_{\nu N}(\epsilon)\,\frac{\partial e_{\mathrm{eq},\epsilon}}{\partial T}.
\end{equation}
Likewise, we can express the right-hand side of Eq.~\eqref{eq:Diffusive-Flux} as $\partial e_\mathrm{eq}/\partial z=(\partial e_\mathrm{eq}/\partial T)\nabla T$. Comparing the expressions and using the explicit formulas for $e_{\mathrm{eq},\epsilon}$ and $e_\mathrm{eq}$ for a fermions without $\mu$, one finds that the average MFP in Eq.~\eqref{eq:Diffusive-Flux} is
\begin{equation}\label{eq:Rosseland}
    \bar\lambda=\int_0^\infty d\epsilon\,\lambda_{\nu N}(\epsilon)\,
    \frac{30\,\epsilon^4 e^{\epsilon/T}}{7\pi^4\,T^5(e^{\epsilon/T}+1)^2},
\end{equation}
which is the usual fermionic Rosseland average.

After including a chemical potential, one cannot directly define a Rosseland average because two derivatives $\partial e_{\mathrm{eq},\epsilon}/\partial T$ and $\partial e_{\mathrm{eq},\epsilon}/\partial \mu$ appear and one cannot define one common effective MFP that would appear in an equation of the form of Eq.~\eqref{eq:Diffusive-Flux}. These questions were discussed a long time ago, for example, by Bludman and van Riper \cite{Bludman:1978} and van den Horn and Cooperstein \cite{vandenHorn:1986}.

With the goal of explaining why secret neutrino interactions do not fundamentally change diffusive energy transfer we study explicitly only the hypothetical case where the energy flux is driven by a temperature gradient, not a chemical-potential gradient. As mentioned earlier, realistically energy is anyway transported mainly by convection.

Another question is the role of scattering on nucleons instead of absorption. For heavy-flavor neutrinos this is the dominant contribution to the opacity. However, once we have large $\nu$SI, and if these affect all flavors, all of them will form a common neutrino fluid for which the dominant opacity derives from the CC interactions of the electron flavor and the muon flavor (in the presence of muons). Therefore, we will not explicitly worry about scattering.

\subsection{Secretly interacting neutrino fluid}

\label{sec:NonrelSecret}

We now turn to large $\nu$SI, meaning $\lambda_{\nu\nu}\ll\lambda_{\nu N}$, so that neutrinos equilibrate among each other between collisions with the nuclear medium. We should picture neutrinos as a fluid that streams along the temperature gradient with a small bulk velocity $v$, which from dimensional analysis is $\mathcal{O}(\lambda_{\nu N}/r)$. The neutrino fluid maintains LTE with the medium so that, apart from small corrections, it must have the blackbody energy density $e_\mathrm{eq}$ prescribed by the properties of the background, analogous to the standard case. In the medium frame, the neutrino distribution must be thermal with a small anisotropic disturbance that allows it to carry an energy flux. In the fluid frame, the distribution must be isotropic, drifting with a small $v$ relative to the medium. For the fluid description, we adopt relativistic hydrodynamical modeling (see, e.g., Ref.~\cite{Weinberg:1972kfs}); for this section, we will only need the form for nonrelativistic bulk velocities, while the equation of state is relativistic throughout this work.

The exact properties of the neutrino fluid in its rest frame are not fully fixed by our assumptions. While $\nu$SI inevitably allow for number-changing processes, these need not be fast enough to establish chemical equilibrium within the neutrino fluid so that only kinetic equilibrium may obtain. However, for our present consideration this point makes no difference because the medium and fluid frames are the same up to the small drift velocity.

With these insights, it is surprisingly simple to derive the diffusive equation of energy transfer, limiting ourselves to the same model as in the standard case, i.e., neutrinos interact with the medium only by absorption and emission. The key idea is that the relativistic neutrino fluid has a pressure $p=\rho/3$ that must change along the radial direction if the energy density is $\rho(z)=e_\mathrm{eq}(z)$ as prescribed by LTE. 

On the other hand, in a steady flow, the momentum flowing through each surface must be conserved. This means that the pressure must be conserved along the direction of its bulk motion, unless it is balanced by a force. It arises from neutrinos being absorbed and emitted with the rate $\Gamma$, the reduced absorption rate defined earlier, but in an asymmetric manner because of the drift velocity. The momentum carried by each volume element is $(\rho+p)v=4\rho v/3$; notice the appearance here of the enthalpy, rather than the energy density. Therefore, the momentum lost per unit volume per unit time is $-4\Gamma v \rho/3$, which must be balanced against the pressure gradient $\partial p/\partial z$. If $\Gamma=1/\lambda_{\nu N}$ does not depend on energy and with $\rho=e_\mathrm{eq}$, we thus find for the energy flux
\begin{equation}\label{eq:flux-secret}
   F=\frac{4}{3}e_\mathrm{eq} v=-\frac{\lambda_{\nu N}}{3}\,\nabla e_\mathrm{eq},
\end{equation}
of the same form as Eq.~\eqref{eq:Diffusive-Flux}. The drift velocity is 
\begin{equation}\label{eq:diff_velocity}
    v=\frac{\lambda_{\nu N}\,|\nabla \log e_\mathrm{eq}|}{4}\ll1,
\end{equation}
which is small due to the original diffusion limit. It is essentially the ratio between the MFP and the medium's temperature scale height.

Once we know the bulk velocity, we can express explicitly the distribution function of the drifting neutrinos. Since $v\ll 1$, we can use a nonrelativistic expansion in terms of the small velocity; we limit ourselves to the case of vanishing chemical potential. Then the corresponding expression is $f^\mathrm{th}\left[\epsilon(1-v\cos\theta)\right]$, which we Taylor expand~as
\begin{equation}
    f=f^\mathrm{th}-\frac{\partial f^\mathrm{th}}{\partial \epsilon}\epsilon v\cos\theta. 
\end{equation}
We now replace the drift velocity from Eq.~\eqref{eq:diff_velocity}, with $|\nabla \log e_\mathrm{eq}|=4|\nabla \log T|$. Furthermore, we use the identity $\partial_\epsilon f^\mathrm{th} \epsilon=-\partial_T f^\mathrm{th} T$ and obtain
\begin{equation}
    f=f^\mathrm{th}-\frac{\partial f^\mathrm{th}}{\partial z}\lambda_{\nu N}\cos\theta,
\end{equation}
which correctly reproduces the noninteracting result in the gray approximation.

In the fluid case, all neutrino modes interact with all others, so a monochromatic case similar to Eq.~\eqref{eq:monochromatic} cannot be contemplated. What if the interaction with the nucleons is not gray (not independent of energy)? In this case we need the spectral average of the force and thus the spectral average of $\Gamma \rho$ which is $\int d\epsilon\,e_{\mathrm{eq},\epsilon}/\lambda_{\nu N}(\epsilon)$. Notice that the drift velocity is common to all $\epsilon$ modes so that $v$ factors out of this averaging procedure. Therefore, in the non-gray case, we recover an equation of the form of Eq.~\eqref{eq:Diffusive-Flux} with
\begin{eqnarray}\label{eq:Rosseland-Fluid}
    \bar\lambda_{\rm fluid}&=&
    \left[\frac{1}{e_\mathrm{eq}}\,\int d\epsilon\,\frac{e_{\mathrm{eq},\epsilon}}{\lambda_{\nu N}(\epsilon)}\right]^{-1}
    \nonumber\\
    &=&\left[\frac{120}{7\pi^4 T^4}
    \int_0^\infty d\epsilon\,\frac{\epsilon^3}{e^{\epsilon/T}+1}\,\frac{1}{\lambda_{\nu N}} \right]^{-1},
\end{eqnarray}
different from the Rosseland average.

If $\nu$SI affect all flavors, and if we ignore chemical potentials, then they together form the neutrino fluid with $\rho=6e_\mathrm{eq}$ if we think of $e_\mathrm{eq}$ as the blackbody distribution of a single nondegenerate neutrino degree of freedom. Of these, $\nu_e$ and $\bar\nu_e$ (and possibly $\nu_\mu$ and $\bar\nu_\mu$) interact by CCs, affecting the entire fluid, whereas an additional force derives from the weaker NC scattering that is equal for all flavors. However, these depend on scattering angle and thus require a more careful treatment.

In the standard diffusive transport case, most of the energy flux is carried by those flavors that only interact by the smaller NC rates, but which are now slowed down by their indirect participation in the CC interactions. Therefore, the overall diffusive energy flux will be smaller than standard, facilitating the appearance of convection. A detailed treatment would require a flavor model of $\nu$SI. On the other hand, even in the standard case, the possible existence of fast-flavor conversion even deeply inside the PNS imply that there is no ``standard'' case to compare with. 

In any event, a numerical SN simulation that includes $\nu$SI will need to implement these effects, in particular in the decoupling region where the diffusion approximation no longer applies. Our discussion here mainly serves to clarify a number of conceptual points. The main result is that there is no mysterious slowing-down of the energy flux by neutrinos being a fluid. But on the other hand, there are quantitative differences arising from the different spectral average of the interaction rate and from coupling all flavors indirectly to the CC rates.

\section{Energy transfer between two blackbody plates}

\label{sec:plates}

We now turn to the more complex regime in which neutrinos are emitted from the PNS surface. In the standard, noninteracting case, deviations from isotropy are more pronounced, and therefore we can expect differences in the secretly-interacting case. 

As a preliminary exercise, we tackle a simplified stationary setup with neutrinos transferring energy between two blackbodies held at constant temperatures $T_1$ and $T_2$. In the SN problem, there is only one surface, of course, but as we will see, the limit $T_2\to 0$ essentially leads to steady-state single blackbody emission in a vacuum. We begin with plane geometry, with two plates fixed at $T_1$ and $T_2$, and later generalize to spherical geometry. 

\subsection{Setup of the problem}

In our schematic setup, we make the simplest assumptions that capture the essence of what large $\nu$SI would do. As in the previous section, we assume that $\lambda_{\nu\nu}\ll \lambda_{\nu N}$ so that neutrinos act as a fluid on scales defined by the interaction rate with the medium. We assume that emission and absorption are the main interaction channels, that $\lambda_{\nu N}$ does not depend on energy, and that it represents the reduced opacity. Moreover, the medium properties are such that $\nu$ and $\bar\nu$ interact equally, i.e., deeply inside the radiating bodies, neutrinos exist as fermionic blackbody radiation without chemical potential. Therefore, in LTE the energy density is
\begin{equation}
    e_{\mathrm{eq},i}=\frac{7\pi^2 T_i^4}{40},\quad i=1~\mathrm{or}~2.
\end{equation}
Here we have multiplied the standard Fermi-Dirac energy density for vanishing chemical potential by a factor $6$, to account for the six species of neutrinos and antineutrinos.

To fix the notation, we will everywhere denote by $\rho$ the comoving energy density of the fluid which moves with bulk velocity $v$ relative to the medium. In LTE, the radiation energy density defined by the medium properties is $e_\mathrm{eq}$. The fluid energy density in the laboratory frame (defined by the medium at rest) is denoted by $e$ and, by Lorentz transformation, is
\begin{equation}
    e=\frac{4\gamma^2-1}{3}\,\rho=\frac{1+v^2/3}{1-v^2}\,\rho,
\end{equation}
where $\gamma=1/\sqrt{1-v^2}$ is the Lorentz factor.
Furthermore, the fluid energy flux can also be obtained by Lorentz transformation as
\begin{equation}
    F=\frac{4}{3}\gamma^2 v \rho=e\xi,
\end{equation}
where we have introduced a convenient modified velocity variable
\begin{equation}
    \xi=\frac{4\gamma^2 v}{4\gamma^2-1}=
    \frac{4 v}{3+v^2}.
\end{equation}
Notice that $\xi$ varies between $0$ and 1 as $v$ varies between 0 and~1. A special role is played by the speed of sound, $\vs=1/\sqrt{3}$, corresponding to
\begin{equation}
    \xs=\frac{2\sqrt3}{5}\simeq0.693
\end{equation}
for our modified velocity of sound.

The two plates have the same homogeneous density and end abruptly at their surface. They are each held at a constant temperature $T_1$ and $T_2$ throughout, although in Appendix~\ref{sec:Milne} we will also consider a self-consistent temperature profile.

\subsection{Plane geometry}

In our first setup, the two blackbodies are two semi-infinite materials disposed along the $z$ axis and separated by a distance $D$. Despite their extended nature, we will simply refer to them as plates, since neutrinos are really emitted from their surface. As neutrinos behave as a fluid even inside the medium, the boundary conditions must be imposed not at the surface, but rather deep inside. We assume the first surface is located at $z=0$ with temperature $T_1$, and the second one at $z=D$ with temperature $T_2<T_1$. 

The equations for the fluid are the equations of energy and momentum conservation. In the laboratory frame, the energy density is $e$ and the energy flux is $F=e\xi$. Energy flux conservation therefore implies $\partial_t e+\partial_z F=(e_\mathrm{eq}-e)/\lambda_{\nu N}$, where the net rate of energy gain appears on the right-hand side, i.e., the difference between thermal energy gain and loss by collisions with the medium. In a steady state, we drop the time derivative and find
\begin{equation}\label{eq:energy-conservation}
   \partial_z(e\xi)=\frac{e_\mathrm{eq}-e}{\lambda_{\nu N}}. 
\end{equation}
for energy conservation.

As discussed in Sec.~\ref{sec:NonrelSecret} above Eq.~\eqref{eq:flux-secret}, in steady state, the pressure gradient along $z$ must be balanced by the force exerted by the plate absorbing neutrinos. If the neutrino fluid is flowing with a bulk velocity, this constraint generalizes to the law of momentum conservation. If the momentum flux is $\mathcal{M}$, and the momentum density is $F$ (by definition this is equal to the energy flux, so we use the same letter), we thus need to satisfy $\partial_t F + \partial_z \mathcal{M}=-F/\lambda_{\nu N}$, where here no gain term appears because the background medium is isotropic. Notice that the momentum flux, which for nonrelativistic velocities coincides with pressure $p$, has now an additional contribution from the bulk motion. In the neutrino fluid frame, this is $\mathcal{M}=p+(\rho+p)\gamma^2 v^2$, which can be written as \smash{$\mathcal{M}=e(5-2\sqrt{4-3\xi^2})/3$} in our lab-frame variables. Finally, in steady state we need to satisfy $\partial_z\mathcal{M}=-F/\lambda_{\nu N}$ and with $F=e\xi$,
this is
\begin{equation}\label{eq:momentum-conservation}
\partial_z\left(e\,\frac{5-2\sqrt{4-3\xi^2}}{3}\right)=-\frac{e\xi}{\lambda_{\nu N}},
\end{equation}
expressing momentum conservation for our system.

In the central region between the plates ($0<z<D$), the behavior is that of a simple fluid with no energy exchange, corresponding to $\lambda_{\nu N}\to \infty$. In this region, the fluid equations require $e$ and $\xi$ both to be constant, to maintain a constant energy and momentum flux. We will denote their constant values in this region by $e_0$ and $\xi_0$ respectively. 

Inside the left plate, Eqs.~\eqref{eq:energy-conservation} and \eqref{eq:momentum-conservation} must be solved with the condition that for $z\to -\infty$ we have $\xi\to 0$ and $e\to e_{\mathrm{eq},1}$. This condition by itself does not determine a unique solution, and must be complemented by another condition at the right boundary $z=0$. For the moment, we impose the condition $\xi(z=0)=\xi_0$. Later, we will determine $\xi_0$ by a matching to the dynamics in the rest of the space, i.e., we need the second plate to determine a steady-state solution.

As a first step, we determine how the energy density $e$ depends on $\xi$ at every point $z$. To do so, from Eqs.~\eqref{eq:energy-conservation} and
\eqref{eq:momentum-conservation} we explicitly extract $\partial_z\xi$ and $\partial_z e$
\begin{eqnarray}\label{eq:xi-e}
 \partial_z \xi\!&=&\!\frac{(e_{\mathrm{eq},1}-e)(8-6\xi^2-5\sqrt{4-3\xi^2})-3e\xi^2\sqrt{4-3\xi^2}}{e(8-5\sqrt{4-3\xi^2})},
        \nonumber\\[1ex]
 \partial_z e\!&=&\!\frac{3\left[2(e_{\mathrm{eq},1}-e)\xi+e\xi\sqrt{4-3\xi^2}\right]}{8-5\sqrt{4-3\xi^2}}.
\end{eqnarray}
We then take the ratio among these expressions to obtain
\begin{eqnarray}\label{eq:dedxi}
\frac{de}{d\xi}=
\frac{3e\xi\left[2e_{\mathrm{eq},1}+e\left(\sqrt{4-3\xi^2}-2\right)\right]}{f(\xi)e_{\mathrm{eq},1}-g(\xi)e},
\end{eqnarray}
where 
\begin{eqnarray}
    f(\xi)&=&8-6\xi^2-5\sqrt{4-3\xi^2},
    \nonumber\\
    g(\xi)&=&8-5\sqrt{4-3\xi^2}+3\xi^2\left(\sqrt{4-3\xi^2}-2\right).
\end{eqnarray} 
Since $T_1>T_2$, it is clear that $\xi>0$ because the fluid flows from left to right, so this equation can be solved with the boundary condition deeply inside that $e(\xi=0)=e_{\mathrm{eq},1}$.

\begin{figure}[t]
\includegraphics[width=\columnwidth]{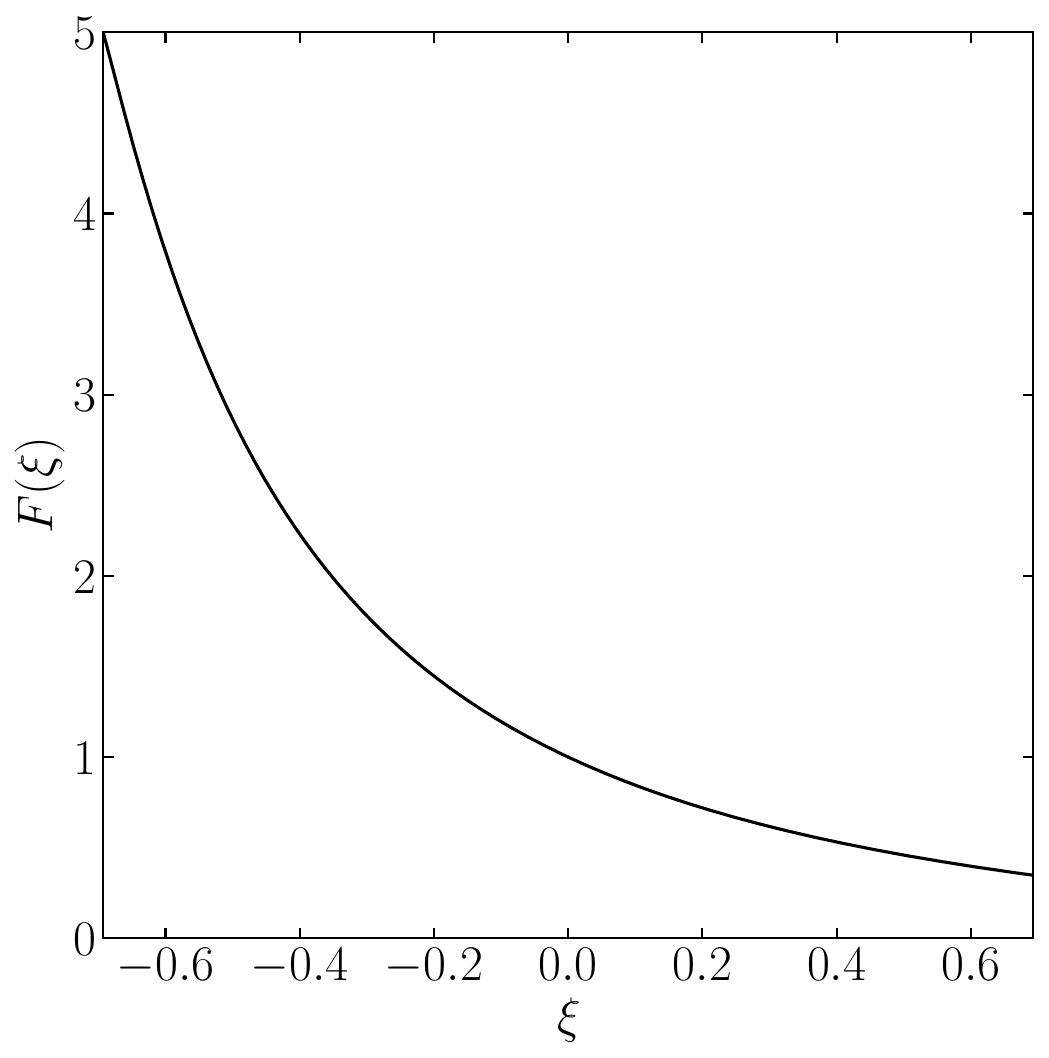}
\vskip-6pt
\caption{Universal function determining the fluid energy density at the plate surface $e_{\rm surface}=e_\mathrm{eq}\,F(\xi)$ as a function of the outflow velocity variable $\xi$, shown in the subsonic range $-\xs<\xi<\xs$. Negative $\xi$ means energy flowing into the plate.}
\label{fig:universal_function}
\end{figure}

In particular, it follows that the energy density and velocity at $z=0$ are related by $e(z=0)=
e_{\mathrm{eq},1} F(\xi_0)$, where $F(\xi)$ is the universal function that solves the equation
\begin{eqnarray}
\frac{dF}{d\xi}= \frac{3F\xi\left[2+F(\sqrt{4-3\xi^2}-2)\right]}{f(\xi)-g(\xi)F},
\quad F(0)=1.
\end{eqnarray}
The solution is shown in Fig.~\ref{fig:universal_function}, limited to the subsonic range $-\xs<\xi<\xs$, which is the only range of physical interest as we will see. Notice that $F(\xi)$ is a decreasing function, so for $\xi>0$, corresponding to an outflow from the body, the energy density decreases as the fluid moves to the right and becomes faster and faster.

For the fluid to the right, the situation is specular; one has the same equation for $de/d\xi$, except that now $\xi$ must be taken with the opposite sign because the flux is ingoing rather than outgoing. Therefore, the energy density and velocity at $z=D$ are related by $e(z=D)=e_{\mathrm{eq},2} F(-\xi_0)$. As the fluid properties between the plates are the same from the perspective of either plate, each characterized by $e_{\mathrm{eq},1,2}$ corresponding to their $T_{1,2}$, the velocity of the fluid between the plates is determined by
\begin{equation}
e_{\mathrm{eq},1} F(\xi_0)=e_{\mathrm{eq},2} F(-\xi_0).
\end{equation}
Evidently, the fluid velocity depends only on the ratio between the temperatures $T_{1,2}$, or equivalently between the blackbody energy densities $e_{\mathrm{eq},1,2}$. From this equation we can deduce both $\xi_0$ and $e_0=e_{\mathrm{eq},1} F(\xi_0)$, the energy density and fluid velocity between the plates.

For $e_{\mathrm{eq},2}/e_{\mathrm{eq},1}$ relatively large (second plate not too cold), the solution is a subsonic outflow from the hot to the cold plate with $\xi_0<\xs$, corresponding to the sound speed. However, when $e_{\mathrm{eq},2}/e_{\mathrm{eq},1}\lesssim 0.07$, the velocity predicted by this method would exceed~$\xs$. However, the outflow from the hot body can never be supersonic. The reason is that, in a steady outflow inside the plates, the energy density rarefies and the velocity grows as a fluid element moves, as long as its speed is subsonic. When the fluid element reaches the sonic point, this tendency inverts, so one can never have a fluid element accelerated to supersonic velocities for a steady outflow. Mathematically, this shows up in the divergence of $\partial_z e$ and $\partial_z \xi$ as $\xi$ reaches $\xs$.

Thus, if $e_{\mathrm{eq},2}/e_{\mathrm{eq},1}\lesssim 0.07$, the outflow velocity from the hot body is fixed to be the speed of sound, \textit{independently of the cold body}. The same conclusion follows from an entirely different argument: once the motion becomes sonic, when $e_{\mathrm{eq},2}/e_{\mathrm{eq},1}\simeq 0.07$, the flow cannot be influenced by a change in the cold body which lies in front of the fluid flowing so that no information can travel toward the hot plate. Therefore, if $T_2$ is lowered further, the flow must remain frozen to its sonic structure.

\begin{figure}
\includegraphics[width=\columnwidth]{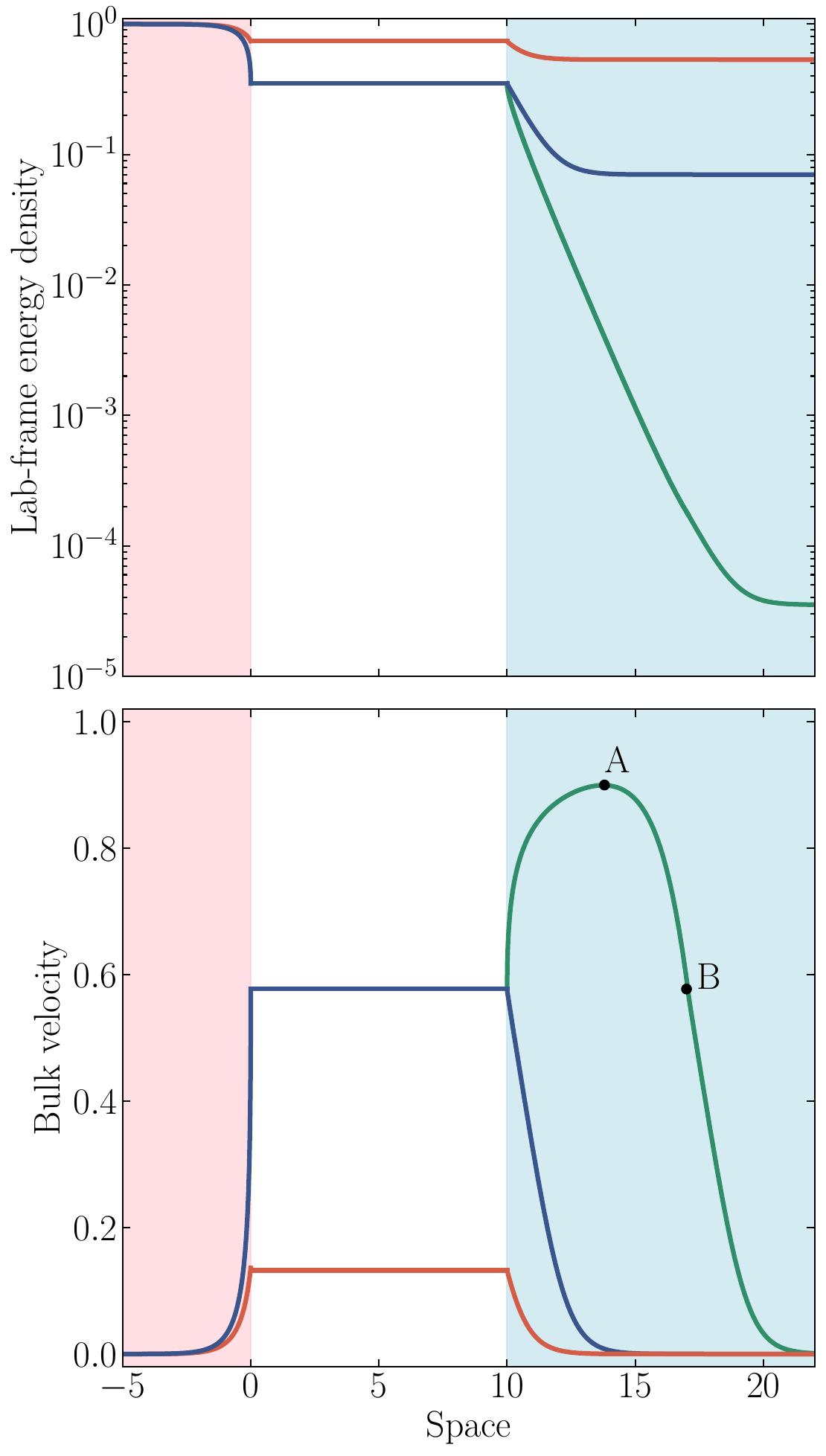}
\caption{Steady-state flow of the neutrino fluid between two plates at temperatures $T_2>T_1$. Energy density in units of $e_{\mathrm{eq},1}$, the blackbody density corresponding to the hot plate $T_1$. We vary $T_2$ and thus $e_{\mathrm{eq},2}$ of the cold plate from subsonic transfer ($0.07\lesssim e_{\mathrm{eq},2}/e_{\mathrm{eq},1}<1$) in red, sonic transfer in blue, and sonic energy transfer with supersonic motion in the cold body ($0\leq e_{\mathrm{eq},2}/e_{\mathrm{eq},1}\lesssim0.07$) in green. In the lower panel, we show the bulk velocity $v$, not $\xi$, with sound speed at $\vs=1/\sqrt{3}\simeq0.577$. We use $D=10$ for the distance between the two surfaces.}\label{fig:steady_two_body}

\end{figure}

These different behaviors are summarized in Fig.~\ref{fig:steady_two_body}, where we show the three regimes of flow: (i)~$e_{\mathrm{eq},2}/e_{\mathrm{eq},1}\gtrsim0.07$ (red line), there is a subsonic flow of neutrinos from the hot to the cold plate, with the energy density damping to the equilibrium values far away from the surfaces in both bodies. (ii)~$e_{\mathrm{eq},2}/e_{\mathrm{eq},1}\simeq0.07$ (blue line), the flow is sonic, and the main properties are the same; this regime corresponds to the maximum energy outflow from the hot plate. (iii)~$e_{\mathrm{eq},2}/e_{\mathrm{eq},1}\lesssim0.07$ (green line), the outflow from the hot plate is the same as (ii) and as large as it can be, with the speed of sound.

Also in this third regime, the energy density decreases to its cold-plate equilibrium value, but the bulk velocity in the cold plate is now quite different. In this regime, from the point of view of the cold body, one has a steady inflow of energy at the speed of sound, so the boundary conditions here are that $\xi(z=D)=\xi_s$, and $e(z=D)\simeq 0.348\,e_{\mathrm{eq},1}>5 e_{\mathrm{eq},2}$. Now, if $e(z=D)=5 e_{\mathrm{eq},2}$ the velocity would monotonically decrease from $\vs$ to zero, while $e$ would drop to its equilibrium value $e_{\mathrm{eq},2}$. However, if $e(z=D)>5 e_{\mathrm{eq},2}$, as is the case for this regime, the tendency is in fact opposite, as one can verify from the equation of motion. The velocity actually \textit{increases} and the motion becomes supersonic inside the cold body, with $e$ decreasing. 

This trend continues until a critical velocity is reached, which is determined by the condition $\partial_z \xi=0$, or, equivalently, by the vanishing of the denominator of Eq.~\eqref{eq:dedxi}, corresponding to the maximum of the green line in Fig.~\ref{fig:steady_two_body} at point~A. Here, the velocity starts to decrease until it reaches the speed of sound again (point~B), where $e(\xi=\xs)=5 e_{\mathrm{eq},2}$. One way to prove this result is to integrate $d\xi/de$ from Eq.~\eqref{eq:dedxi}\footnote{One cannot integrate the equation directly to obtain $e(\xi)$ since in this scenario this is a non-monotonic function; the energy density decreases as the velocity first increases and then decreases.} and to notice that for $\xi=\xs$, the derivative $d\xi/de$ becomes infinite at $e=5 e_{\mathrm{eq},2}$, so independently of the initial value $e(z=D)$ the energy density profile will reach $e=5e_{\mathrm{eq},2}$ when it passes again through the sonic point~B. But here the profile has just the correct value of energy density for the velocity to smoothly go to $0$ as the energy density smoothly goes to $e_{\mathrm{eq},2}$ as $z\to +\infty$. 

Notice that the velocity behavior in the cold plate is discontinuous as we pass between the regimes: the slope jumps from negative to positive between the blue and green line at the surface of the cold plate (Fig.~\ref{fig:steady_two_body}). In fact, the slope of the green line is infinite at the surface as one can glean from the denominator of Eq.~\eqref{eq:xi-e} diverging when $\xi\to\xs$.

Notice also that as $T_2\to 0$, point A, where the velocity in the cold body is largest, moves towards $z\to \infty$, and in turn the maximum velocity becomes closer to the speed of light. At zero temperature, the fluid inside the cold body moves with an ever-increasing velocity to the right while its density drops to zero.

\begin{figure}
    \includegraphics[width=\columnwidth]{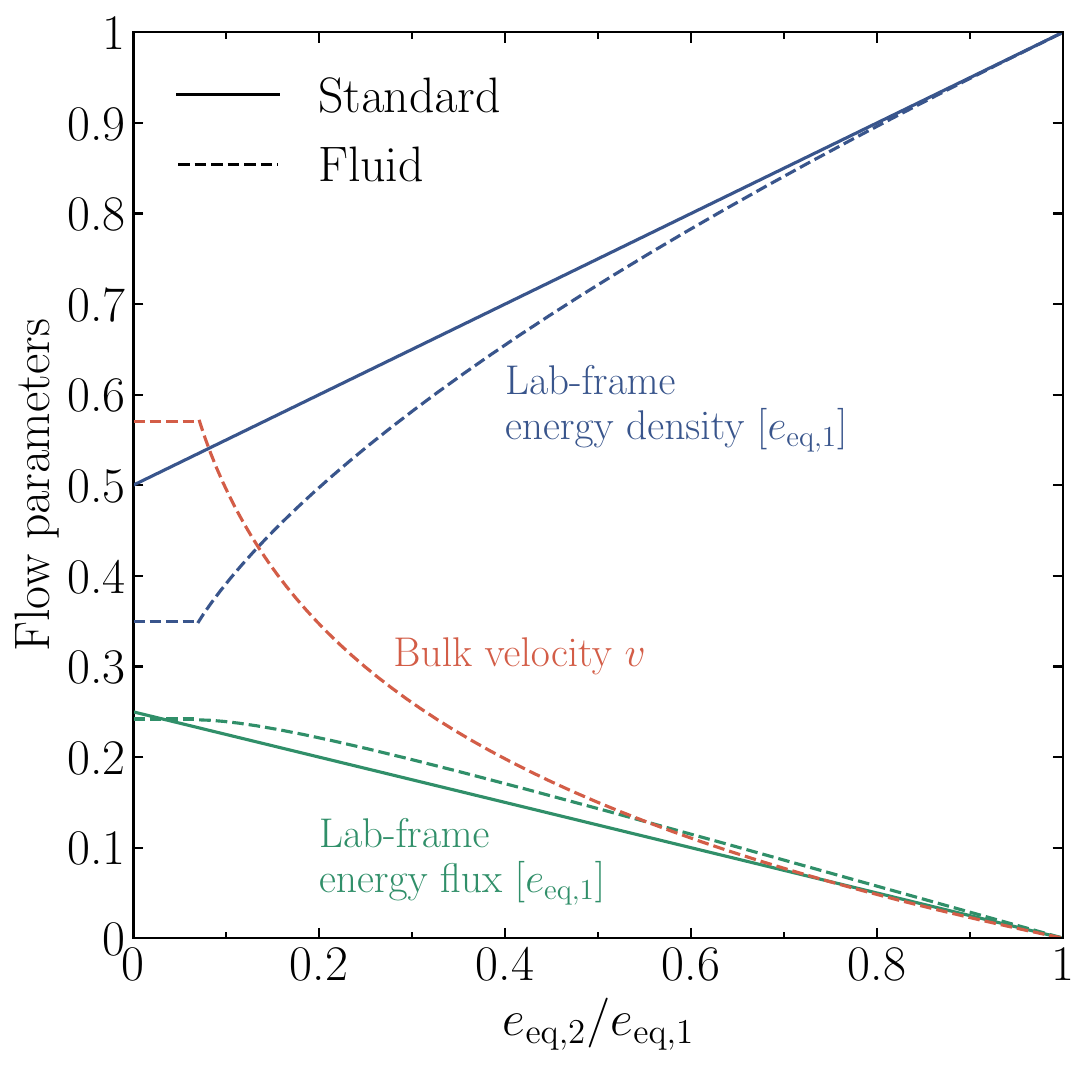}
    \caption{Parameters of the flow between the two blackbodies as a function of $e_{\mathrm{eq},2}/e_{\mathrm{eq},1}=(T_2/T_1)^4$. {\em Solid lines:} Standard radiation according to the Stefan-Boltzmann Law. {\em Dashed lines:} Fluid with large $\nu$SI. We show the bulk velocity (only for the fluid) as well as the energy density and energy flux in the lab frame in units of $B_1$. }\label{fig:influx_no_fluid}
\end{figure}

The flow parameters are also shown in Fig.~\ref{fig:influx_no_fluid} as a function of $e_{\mathrm{eq},2}/e_{\mathrm{eq},1}=(T_2/T_1)^4$. Solid lines refer to standard radiation, where the energy density between the plates is trivially $(e_{\mathrm{eq},1}+e_{\mathrm{eq},2})/2$ and the energy flux $(e_{\mathrm{eq},1}-e_{\mathrm{eq},2})/4$. Dashed lines refer to the self-interacting neutrino fluid. In both cases, the energy flux between the two plates is very similar. In particular, for vanishing $T_2$, the energy flux is $0.24\,e_{\mathrm{eq},1}$ in the fluid case, compared with 0.25 in the standard case. On the other hand, the energy density exhibits a noticeable drop in the fluid case and is smallest in the limit of sonic outflow, for $e_{\mathrm{eq},2}/e_{\mathrm{eq},1}\lesssim0.07$, where $e_0=0.35\,e_{\mathrm{eq},1}$, to be compared with 0.50 for the standard Stefan-Boltzmann case.

Our solution for sonic outflow is similar to the one found in Appendix~H of Ref.~\cite{Chang:2022aas}, although these authors made different physical assumptions. They set energy exchange between the neutrino fluid and the nuclear medium to zero, whereas we assumed LTE deep inside the plate, so the properties of the neutrino fluid are fixed by the nuclear medium. On the other hand, physically in the outer PNS layers (the atmosphere), in steady state indeed there will be no energy exchange. However, such a situation requires a self-consistent temperature profile along the lines of our Appendix~\ref{sec:Milne}. Our assumption of an isothermal plate with externally fixed properties is of course also unphysical because one needs to assume an unspecified mechanism that keeps the plate isothermal despite radiating. 

While sonic outflow is a generic feature found both in Ref.~\cite{Chang:2022aas} and our treatment, our approach has the advantage of relating the properties of the escaping fluid to the properties (the temperature) of the source.

\subsection{Number conservation} 

\label{sec:number_conservation}

So far, our results have only used the fact that $\nu$SI turn the neutrino gas into a fluid, and thus we have used energy and momentum conservation and the ultra-relativistic equation of state $p=\rho/3$. A separate question is the internal state of the neutrino fluid which, by assumption, relaxes to kinetic equilibrium. If it also relaxes to chemical equilibrium depends on the speed of number-changing processes such as $\nu\bar{\nu}\to\nu\bar{\nu}\nu\bar{\nu}$. In this case, the internal fluid temperature follows from the energy density in its rest frame. We here always model the background medium as acting symmetrically for $\nu$ and $\bar\nu$, otherwise the neutrino fluid can also inherit an excess lepton number before decoupling.

If number-changing processes are too slow to be relevant, the neutrino number density must be determined by explicitly integrating an equation of number conservation. We denote by $n$ the comoving neutrino number density, so that the equation of number conservation, assuming again an energy-independent neutrino nucleon mean free path, is
\begin{equation}\label{eq:number_conservation}
    \partial_z (\gamma n v)=\frac{n^\mathrm{th}-\gamma n}{\lambda_{\nu N}},
\end{equation}
where $n^\mathrm{th}$ is the equilibrium neutrino number density within the radiating body. Notice that the velocity profile is entirely determined by energy and momentum conservation, so this new equation is not a novel dynamical equation, but must only be solved within the fixed velocity profile that we have determined earlier. 

For $T_2=0$, which is the case of the hot plate radiating into vacuum, we have integrated Eq.~\eqref{eq:number_conservation} numerically assuming $n=n^\mathrm{th}$ for $z\to-\infty$, and using the velocity profile that solves Eqs.~\eqref{eq:energy-conservation} and \eqref{eq:momentum-conservation}. We find that at the sonic point, namely at the surface of emission, the neutrino number density $n=0.29\,n^\mathrm{th}$. Therefore, in the case of number-conserving dynamics, this is the necessary boundary condition to describe the dynamics outside of the radiating surface.

\subsection{Spherical geometry}

\label{sec:spheres}

We now extend these results to spherical geometry. In place of the two plates we use an internal, hot sphere of radius $r_1$ and temperature $T_1$ and an external, cold sphere of radius $r_2$ and temperature $T_2<T_1$. We will later also comment on the opposite case $T_2>T_1$. Assuming that $\lambda_{\nu N}\ll r_1<r_2$, we can use plane-parallel geometry within the spheres themselves. The only difference is the geometric behavior in the vacuum gap between the surfaces, where the energy density and the bulk velocity were constant. Instead, in steady state, the spherical equations must now hold
\begin{eqnarray}
\kern-1em&&\partial_r\left(e\xi r^2\right)=0, \\ \nonumber 
\kern-1em&&\frac{1}{3r^2}\partial_r\left[e\left(5-2\sqrt{4-3\xi^2}\right) r^2\right]-\frac{2e}{3r}(\sqrt{4-3\xi^2}-1)=0.
\end{eqnarray}
These equations admit an implicit solution
\begin{equation}\label{eq:steady_velocity_profile}
G(\xi)=\frac{\left(2+\sqrt{4-3\xi^2}\right)^3}{324\left(\!\sqrt{4-3\xi^2}-1\right)^2\!\left(2-\sqrt{4-3\xi^2}\right)}
=\left(\frac{r}{\rs}\right)^4.
\end{equation}
We choose the integration constant $\rs$ as a sonic radius at which the velocity equals the speed of sound. 

\begin{figure}
    \centering
    \includegraphics[width=\columnwidth]{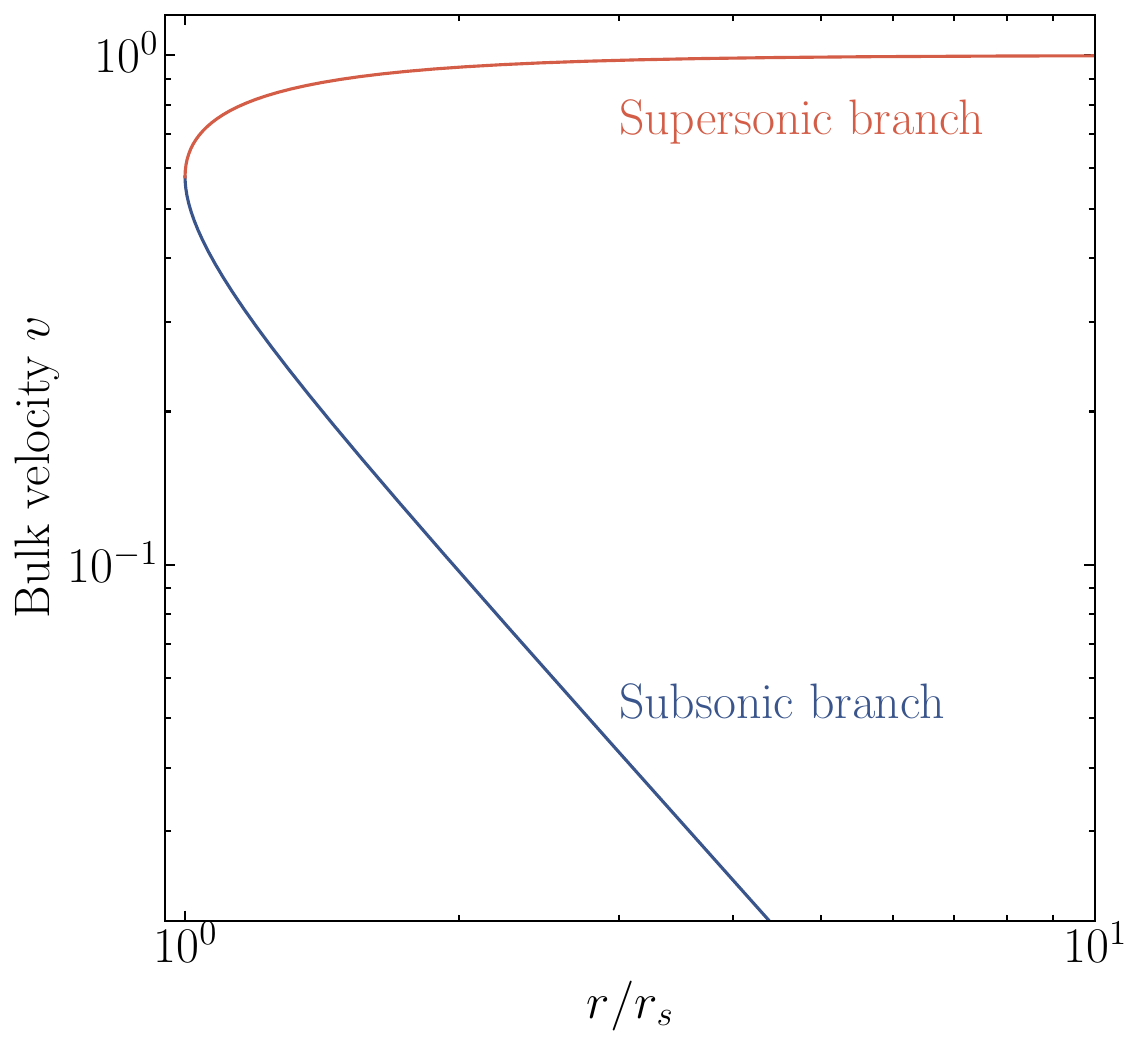}
    \caption{Velocity profile for spherical geometry according to Eq.~\eqref{eq:velocityprofile}, measured from the sonic point at $r=\rs$.}
    \label{fig:radialmotion}
\end{figure}

This solution admits an even simpler form in terms of the original velocity variable
\begin{equation}\label{eq:velocityprofile}
    v(1-v^2)=\frac{2}{3\sqrt{3}}\left(\frac{\rs}{r}\right)^2.
\end{equation}
The velocity profile is shown in Fig.~\ref{fig:radialmotion}. There are two branches for $v(r)$, corresponding to two physically different behaviors. If the velocity at the inner surface $r_1$ is subsonic, then it tends to diminish with increasing $r$. On the contrary, if the velocity at $r_1$ is supersonic, it tends to increase to the speed of light. It is a fundamental conclusion that a steady subsonic flow decelerates in the direction of propagation, while a supersonic flow accelerates. Notice however that if the flow is inward -- if the outer sphere is hotter than the inner one -- the two cases interchange, and a subsonic flow would accelerate in the inward direction of propagation.

The supersonic flow shown here corresponds to the steady wind solution obtained in Ref.~\cite{Chang:2022aas} for the motion outside of the proto-neutron star. On the other hand, the completion of the solution below the blackbody surface, which in Ref.~\cite{Chang:2022aas} corresponds to the PNS, is in our case quite different, since we consider that the neutrino fluid is kept in thermal equilibrium inside the blackbody.

At radius $r_1$, escaping the hot sphere, the unknown escape velocity is $\xi_1$. Entering the cold sphere, at radius $r_2$, the corresponding $\xi_2$ is determined by the implicit equation
\begin{equation}\label{eq:final_velocity}
\frac{G(\xi_2)}{G(\xi_1)}=\left(\frac{r_2}{r_1}\right)^4.
\end{equation}
If $\xi_1$ is subsonic, one must choose the branch where both $\xi_1$ and $\xi_2$ are subsonic. As we know from our previous plane case, for small enough $e_{\mathrm{eq},2}/e_{\mathrm{eq},1}$, the outflow velocity $\xi_1$ will become sonic, in which case one must take the supersonic branch $\xi_2>\xi_1$. In both cases, we can define a function $\xi_2(\xi_1, r_2/r_1)$ giving the velocity at the outer sphere as a function of the inner velocity and the ratio of the radii. We can now construct an explicit steady spherical solution. As in plane geometry, the behavior of the solution depends on whether the outflow velocity from the inner surface is subsonic or sonic.  

In the subsonic branch, the outflow velocity is entirely determined by the matching of the energy fluxes at the inner and outer sphere. Both $\xi_1$ and $\xi_2(\xi_1,r_2/r_1)$, the latter determined from Eq.~\eqref{eq:final_velocity}, are subsonic. From our discussion in the plane case, we know that the energy density at the emission from the inner sphere is $e_{\mathrm{eq},1} F(\xi_1)$, while the energy density entering the outer sphere is $e_{\mathrm{eq},2} F\left[-\xi_2(\xi_1,r_2/r_1)\right]$. Since the energy fluxes must be equal, we obtain a single equation for the outflow velocity $\xi_1$, which is
\begin{equation}
e_{\mathrm{eq},1} F(\xi_1) \xi_1 r_1^2=e_{\mathrm{eq},2} F\left[-\xi_2(\xi_1,r_2/r_1)\right]\xi_2(\xi_1,r_2/r_1) r_2^2.
\end{equation}
Notice that now the outflow velocity $\xi_1$ depends not only on
$e_{\mathrm{eq},2}/e_{\mathrm{eq},1}$ as in the plane case, but also on $r_2/r_1$. If the predicted outflow velocity exceeds the speed of sound, we return to the earlier argument that it must be exactly sonic. However, in our spherical geometry, the velocity becomes supersonic already in the region between the spheres, and the fluid enters the cold, outer sphere already supersonically. As in the plane case, the flow structure then adjusts itself inside the cold body in such a way as to match the boundary condition at $r\to +\infty$ that $e\to e_{\mathrm{eq},2}$.

\begin{figure}
    \centering
    \includegraphics[width=\columnwidth]{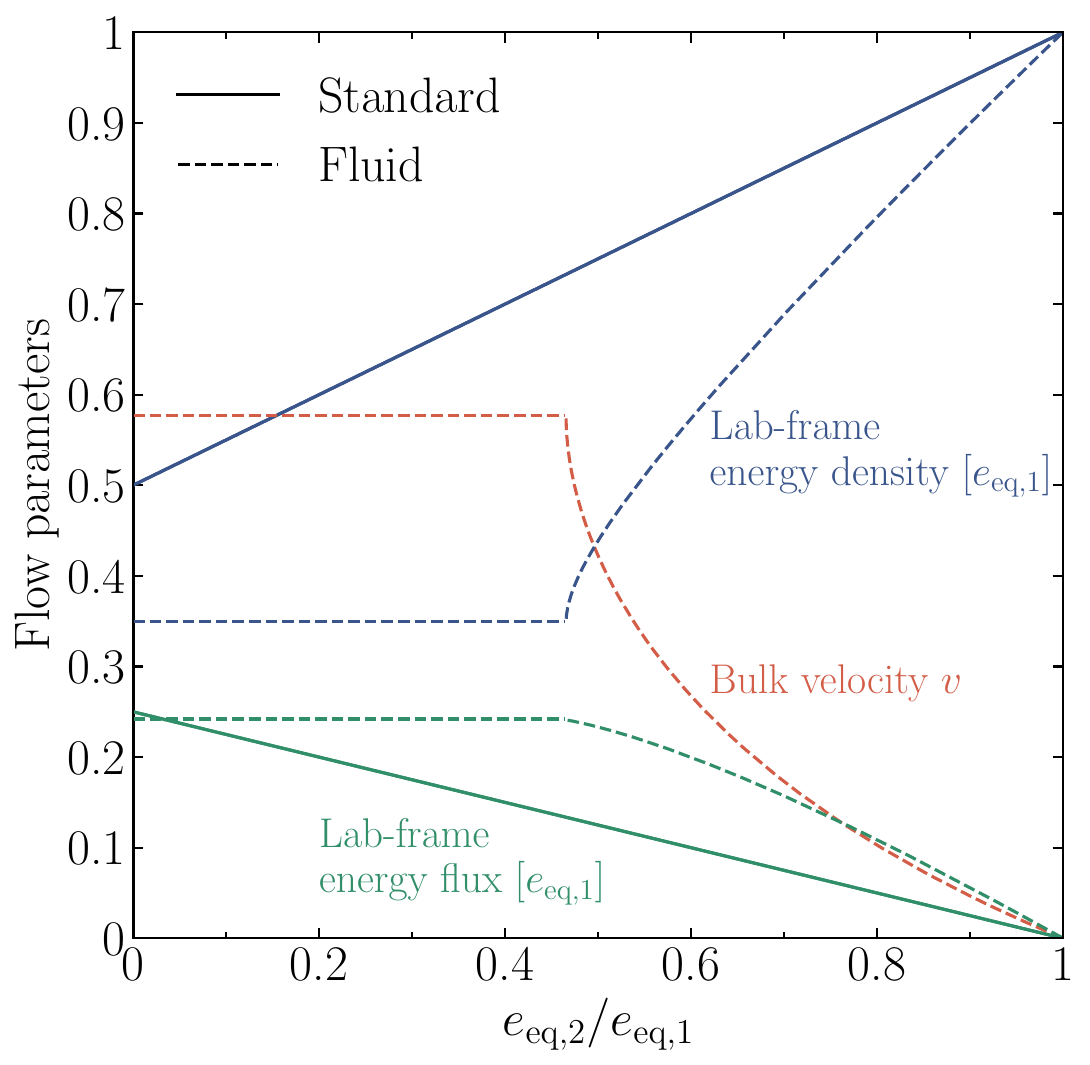}
    \caption{Parameters of the flow between two spherical surfaces
    as a function of $e_{\mathrm{eq},2}/e_{\mathrm{eq},1}=(T_2/T_1)^4$, measured immediately outside the inner sphere. In contrast to plane geometry, these parameters evolve along the radial direction. {\em Solid lines:} Standard radiation. {\em Dashed lines:} Fluid with large $\nu$SI. We show the bulk velocity (only for the fluid) as well as the energy density and energy flux in the lab frame in units of $e_{\mathrm{eq},1}$. We choose a benchmark value for $r_2/r_1=10$.}
    \label{fig:inflow_spherical}
\end{figure}

We show the spherical flow parameters in Fig.~\ref{fig:inflow_spherical} as a function of $e_{\mathrm{eq},2}/e_{\mathrm{eq},1}$ (dashed lines) to be compared with the energy density and flux in the noninteracting case (solid lines), where the energy density at the inner surface would be the average of $e_{\mathrm{eq},1}$ and $e_{\mathrm{eq},2}$. On the other hand, in the $\nu$SI case, the neutrinos from the hot and the cold spheres thermalize and, due to the geometrical dominance of the cold sphere, the energy density at $r_1$ is actually much closer to $e_{\mathrm{eq},2}$. However, the energy flux from the inner to the outer sphere is not very different from the noninteracting case. Especially for $e_{\mathrm{eq},2}/e_{\mathrm{eq},1}\to 0$ (zero-temperature outer sphere), the standard and fluid cases lead to very similar predictions. At a critical ratio $e_{\mathrm{eq},2}/e_{\mathrm{eq},1}\simeq 0.45$, the outflow from the inner sphere becomes sonic and is not anymore influenced by the cold outer sphere. The specific number $0.45$ of course depends on the chosen benchmark value of $r_2/r_1=10$, and slightly changes for other $r_2/r_1$ values.

If the outer sphere is hotter than the inner one, the flow is reverted from outside in. However, the velocity now \textit{increases} in the inward direction until the velocity at the inner sphere becomes equal to the speed of sound. As one further lowers $e_{\mathrm{eq},2}/e_{\mathrm{eq},1}$, the flow cannot change any further, because the velocity at the outer sphere is forced to be subsonic -- it is still true that the outflow from a hot body cannot be supersonic, as we showed in the plane case. Therefore, the flux remains frozen and is now entirely determined by the \textit{cold} body, contrarily to all earlier cases where instead the hot body determined the flow. While this may sound somewhat counterintuitive, there is no contradiction: the flow is now subsonic, and therefore it can be determined by the central cold body. In fact, for $r_2\to\infty$, the cold body is essentially left inside a thermal hotter environment, and it is clear that it must be the cold body which determines the properties of the flow in this regime.

\section{Dynamical Expansion}

\label{sec:dynamical}

The steady state of a blackbody radiating a neutrino fluid into a vacuum can be seen as the previous two-plate solution for a zero-temperature cold plate ($T_2\to 0$). This solution corresponds to a steady, sonic outflow into space, but how is it dynamically reached from a given initial condition? Here we explore this question, assuming the radiation process starts suddenly, mimicking SN core collapse, so initially no neutrinos exist outside. The sudden beginning leads to a front with a weak discontinuity in energy density expanding into space, which gradually approaches a steady outflow near the surface. Once more, we separately discuss plane and spherical geometry. 

We make the same physical assumptions as in the two-plate problem
(only emission-absorption, energy independence of the neutrino MFP, no chemical potential, isothermal radiating body) and denote the equilibrium neutrino blackbody energy density prescribed by the blackbody temperature $T$ as $e_\mathrm{eq}\propto T^4$.

\begin{figure*}
    \centering
    \includegraphics[width=\columnwidth]{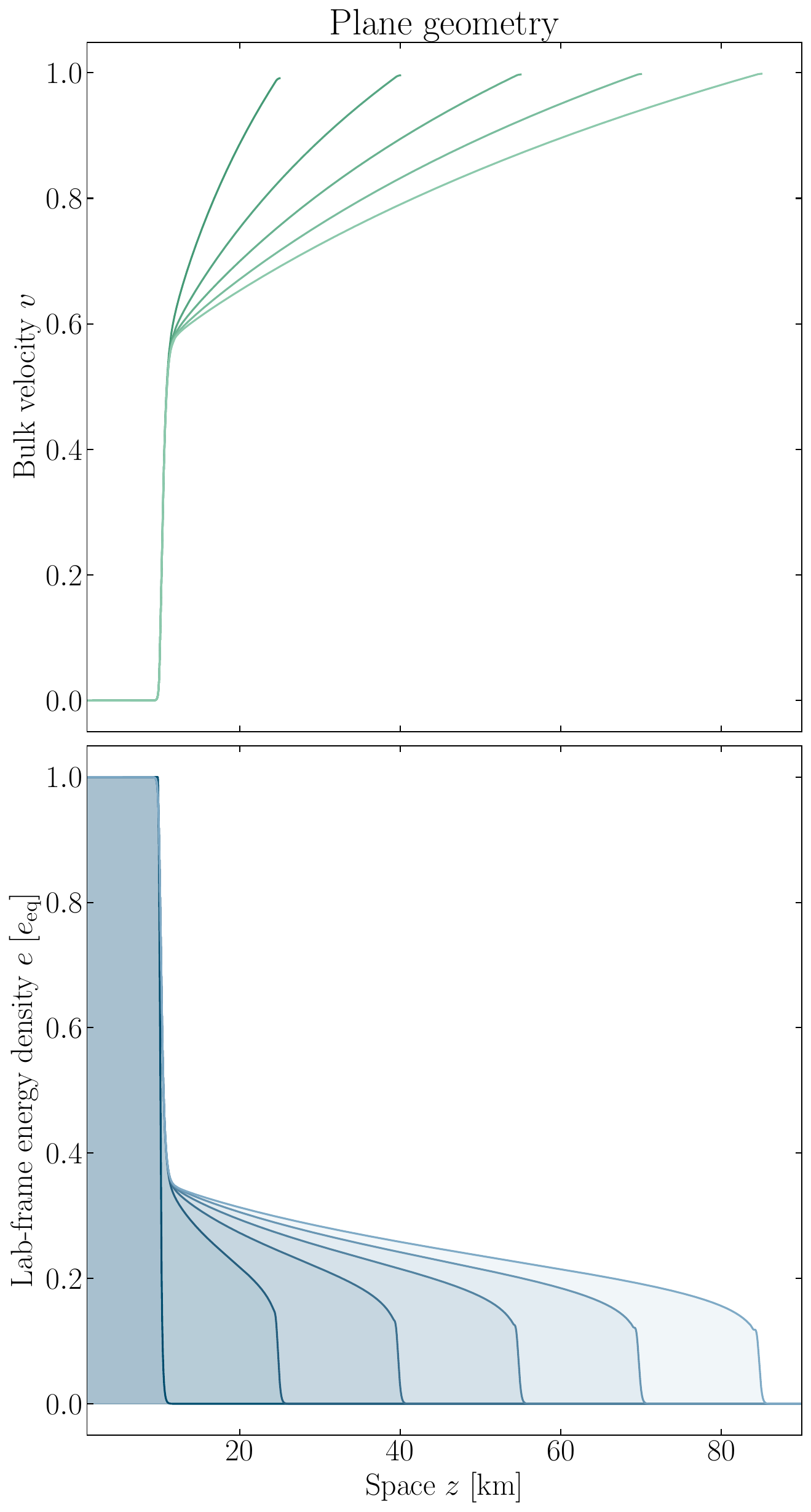}
    \includegraphics[width=\columnwidth]{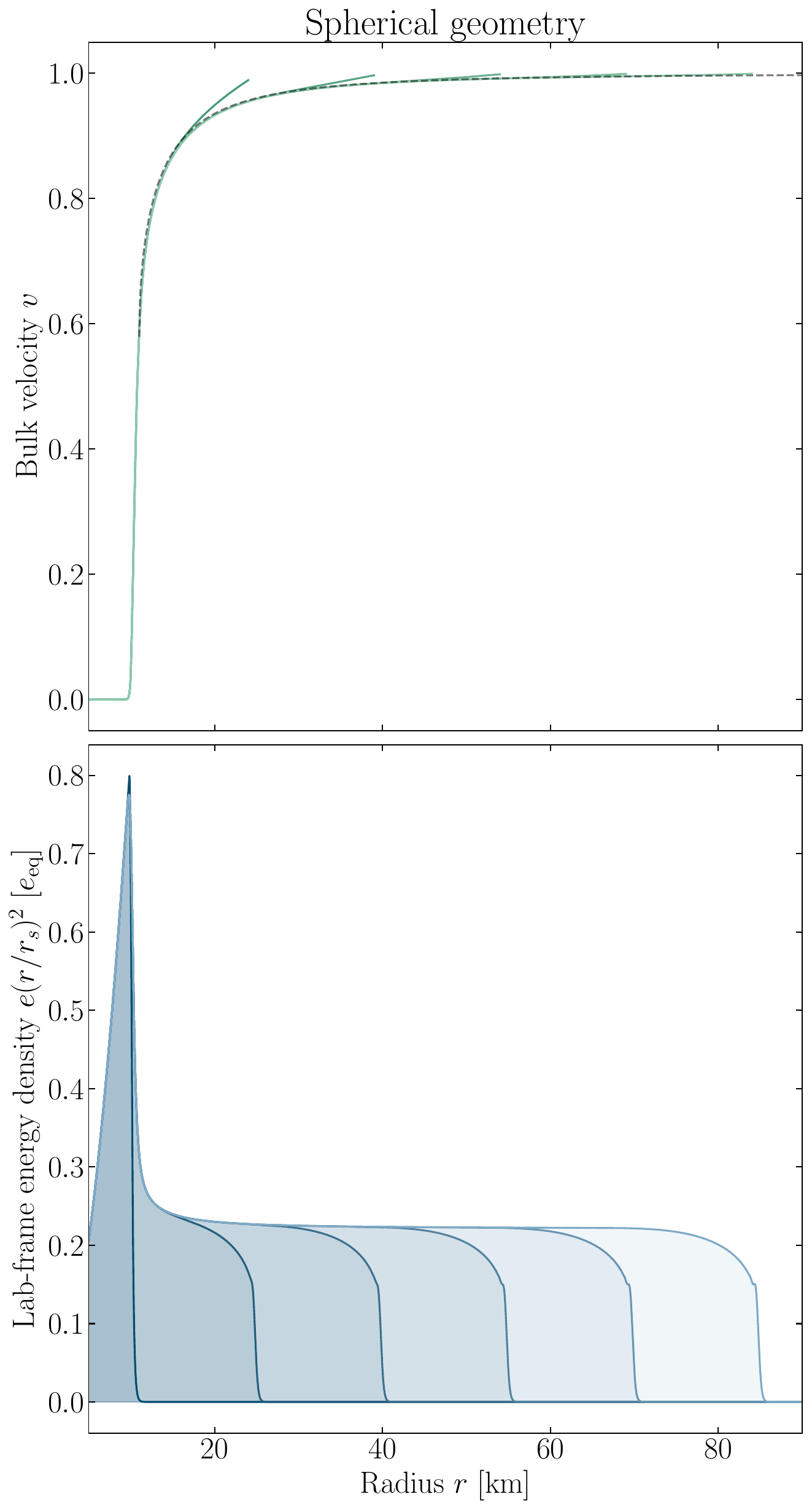}
    \caption{Time evolution of the initial energy density and velocity profile in plane geometry (left) and spherical geometry (right). The progressively lighter shadings correspond to snapshots in time taken every $50$~$\mu$s from $0$~$\mu$s to $250$~$\mu$s. The thin dashed line in the upper right panel denotes the steady velocity profile of Eq.~\eqref{eq:steady_velocity_profile}.}
    \label{fig:snapshots}
\end{figure*}

\subsection{Plane geometry}

In the case of plane geometry, we need to solve the time-dependent fluid equations of energy and momentum conservation
\begin{subequations}
\begin{eqnarray}\label{eq:plane_time_equation}
\kern-2em&&\partial_t e +\partial_z(e\xi)=\frac{e_\mathrm{eq}-e}{\lambda_{\nu N}},
\\[1ex]
\kern-2em&& \partial_t(e\xi)
+\partial_z\left[\frac{e\left(5-2\sqrt{4-3\xi^2}\right)}{3}\right]
=-\frac{e\xi}{\lambda_{\nu N}}
\end{eqnarray}    
\end{subequations}
in analogy to the earlier stationary case of Eqs.~\eqref{eq:energy-conservation} and \eqref{eq:momentum-conservation}. Once more, $e$ is the lab-frame fluid energy density and $F=e\xi$ its energy flux.

We first solve these equations numerically. To avoid numerical instabilities, we use a smoothed surface and modulate the MFP as
\begin{equation}
    \lambda_{\nu N}^{-1}=5~{\rm km}^{-1}\times
    \begin{cases}
        1&{\rm for}~z\leq 10~{\rm km},\\[1ex]
        e^{-6\,\left(z/\mathrm{km}-10\right)}&{\rm for}~z> 10~{\rm km}.
    \end{cases}
\end{equation}
Likewise, as an initial condition  $e(z,t=0)$ for the fluid energy density we use the same profile with the overall coefficient $e_\mathrm{eq}$.

Figure~\ref{fig:snapshots} (left) shows the time evolution that we find numerically. The energy density inside the blackbody nearly immediately evolves to $0.35\,e_\mathrm{eq}$, the value we predicted for the steady sonic outflow in the earlier two-plate model. The neutrino fluid first escapes with a front moving with the speed of light. A rarefaction wave smoothly connects this motion with the speed of sound at the blackbody surface. As the front progressively moves far away, the solution evolves indeed towards the steady state we derived in Sec.~\ref{sec:plates}, with a sonic outflow with constant velocity and constant energy density $e_0=0.35\,e_\mathrm{eq}$. Notice that, since the outflow is sonic, this implies that the comoving energy density at the emission surface is $\rho_0=0.21\,e_\mathrm{eq}$.

In fact, the solution outside of the surface exactly coincide with the one found in Ref.~\cite{Dicus:1988jh} for the complementary problem of a gas inside a container whose piston is suddenly removed. In Ref.~\cite{Dicus:1988jh} this solution was obtained by the method of characteristics, somewhat obscuring its mathematical simplicity. In reality, outside of the blackbody surface, the solution is self-similar and can be expressed in the simple form
\begin{equation}\label{eq:self_similar_velocity}
    v(z,t)=\frac{\sqrt{3}\,(z-\zs)+t}{\sqrt{3}\,t+(z-\zs)},
\end{equation}
and for the comoving energy density
\begin{equation}\label{eq:self_similar_density}
    \rho(z,t)=0.21 \left[\frac{t+z-\zs}{t-z+\zs}\right]^{-2/\sqrt{3}},
\end{equation}
where $\zs$ is the coordinate of the blackbody surface. The factor $0.21$ is chosen to match the boundary condition $\rho=0.21\,e_\mathrm{eq}$ at the blackbody surface for sonic outflow. Since we have taken a profile for $\lambda_{\nu N}$ which is smoothened for numerical reasons, the value of $\zs$ is not exactly equal to $10$~km, as is also visible from Fig.~\ref{fig:snapshots}. However, we have explicitly verified that with a value $\zs\simeq 11$~km the self-similar solution exactly matches our numerical solution at all times.

\subsection{Spherical geometry}

For spherical geometry, the fluid equations to be solved are
\begin{subequations}
\begin{eqnarray}\label{eq:sphere_time_equation}
\kern-4em&&\partial_t e +\frac{\partial_r(e\xi r^2)}{r^2}=\frac{e_\mathrm{eq}-e}{\lambda_{\nu N}},
\\  \nonumber
\kern-4em&& \partial_t(e\xi)+\frac{1}{3r^2}\partial_r\left[e\left(5-2\sqrt{4-3\xi^2}\right) r^2\right]\\
\kern-4em&&\kern7em{}-\frac{2e}{3r}(\sqrt{4-3\xi^2}-1)=-\frac{e\xi}{\lambda_{\nu N}}.
\end{eqnarray}    
\end{subequations}
We solve these equations with the same conditions as for plane geometry, except that now the $z$ coordinate is interpreted as radius~$r$. We show the numerical solution to these equations in Fig.~\ref{fig:snapshots} (right panels). Notice that, in place of the energy density, we here show the combination $e(r/\rs)^2$, where $\rs=11$~km is the approximate radius of the emitting sphere.

Initially, the solution behaves similarly to the plane case, as one can expect. A neutrino weak discontinuity is launched from the surface, and the energy density is here equal to $0.35\,e_\mathrm{eq}$, as we have repeatedly found. As the front moves away, the self-similar profile breaks down; in fact, there is no reason why we should expect a self-similar solution for spherical geometry, where there is a characteristic length, the radius of the emitting surface $\rs$. Instead, the velocity profile rapidly evolves into the steady supersonic solution described implicitly in Eq.~\eqref{eq:steady_velocity_profile}. Thus, in spherical geometry, both the front and the region behind it move with a velocity close to the speed of light, maintaining their shape, with the energy density decreasing as $e\propto r^{-2}$.

\section{End of the signal}

\label{sec:End}

If the emission from the blackbody is interrupted, for example by decreasing the temperature, the corresponding neutrino emission drops. The fluid in the tail of the emission will therefore propagate with a lack of pressure behind it, which may slow it down and ultimately stretch the signal duration. We now examine this question, separately for plane and spherical geometry, and find that in the latter case, the neutrino burst propagates as a shell of constant thickness, in agreement with the usual fireball theory.

\subsection{Plane geometry}

We consider the same setup adopted in Sec.~\ref{sec:dynamical}, except that now after $t_0=20$~km$/c$ (we restore temporarily the speed of light $c$ to emphasize the units) we consider the blackbody spectrum decaying exponentially in time as
\begin{equation}
    e_\mathrm{eq}(t)=e_\mathrm{eq} \exp\left[-0.05\,\frac{c\,(t-t_0)}{\mathrm{km}}\right].
\end{equation}
This form is meant to simulate the turning off of the blackbody, which however happens only with a slow exponential rather than abruptly to avoid dealing with sharp features in the numerical solution.

\begin{figure*}
    \includegraphics[width=\columnwidth]{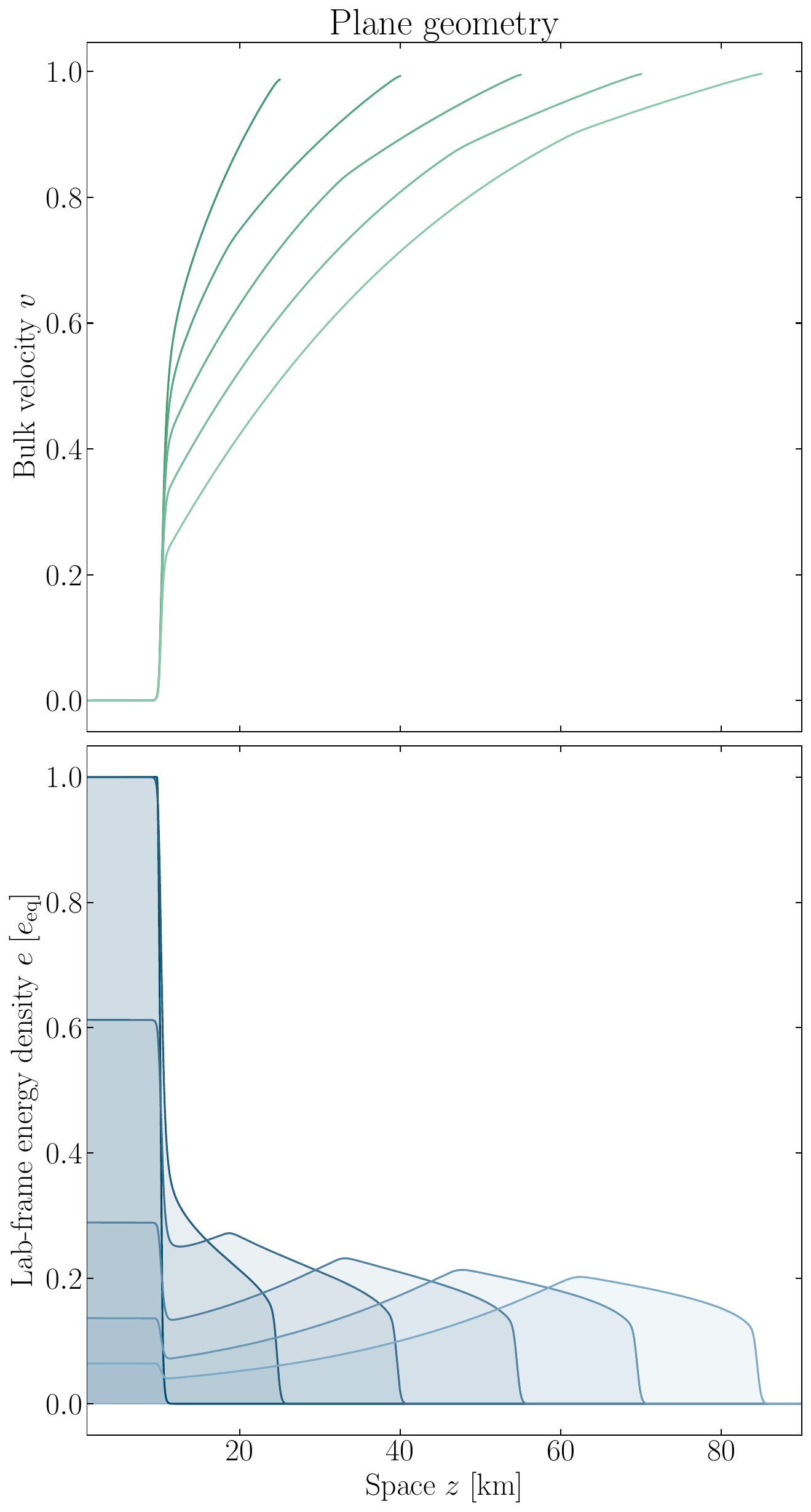}
    \includegraphics[width=\columnwidth]{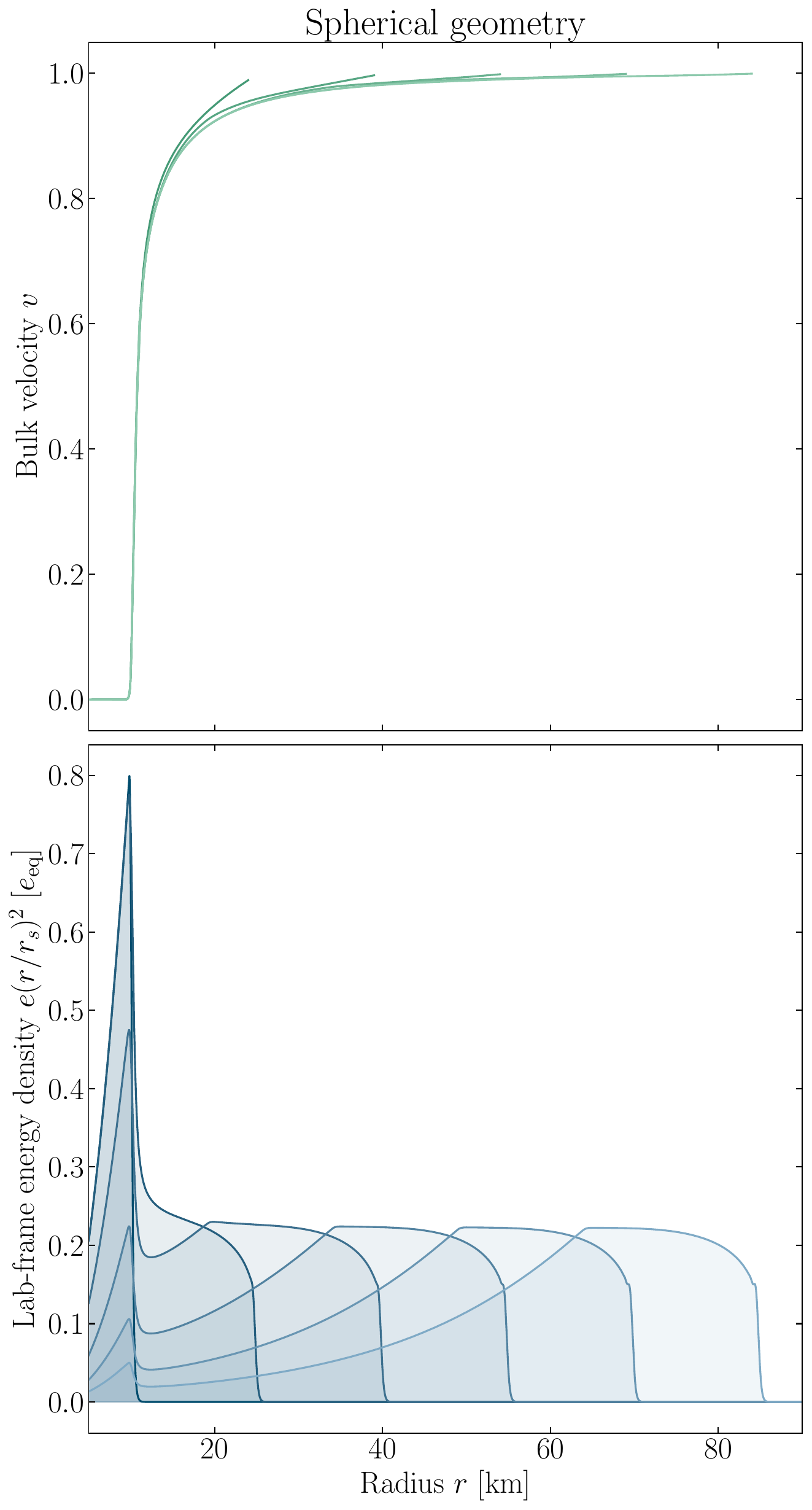}
    \caption{Time evolution of the initial energy density and velocity profile in plane geometry (left) and spherical geometry (right), with an exponential turn-off of the blackbody emission after $t_0=20$~km/$c$. The progressively lighter shadings correspond to snapshots in time taken every $50$~$\mu$s from $0$~$\mu$s to $250$~$\mu$s.}\label{fig:snapshots_td}
\end{figure*}

Figure~\ref{fig:snapshots_td} (left panels) shows the temporal snapshots of the corresponding solution. The region close to the front keeps behaving according to the self-similar solution, unaware that the blackbody has turned off. On the other hand, after the blackbody temperature is dropped, the escape from the emitting surface becomes subsonic, since the fluid inside the blackbody has a lower pressure support and slows down the fluid in front of it. For much later times, the velocity profile would become even negative in the tail, corresponding to the rarefaction wave caused by the lack of pressure; in other words, there is fallback of a small fraction of the fluid. However, our numerical solution develops instabilities at such later times, so we only show snapshots in the early phase.

One can prove that there is always a region that proceeds without being affected by the blackbody turn-off. Signals from the emitting surface can only propagate with a maximum velocity given by the composition of the fluid velocity and the speed of sound. We might call this limiting surface the sound horizon, in front of which the flow is unaffected. Therefore, there is a limiting curve given by the equation
\begin{equation}
    \frac{dz}{dt}=\frac{v(z,t)+\vs}{1+v(z,t)\,\vs},
\end{equation}
with the initial condition that $z(t_0)=\zs$, where $\zs$ is the coordinate of the emitting surface. To the right of this limiting curve, the flow remains given by the self-similar solution. Using Eq.~\eqref{eq:self_similar_velocity}, the limiting curve $z_{\ell}(t)$ is found to be given by the implicit expression
\begin{equation}
    \left(\frac{t+z_{\ell}(t)}{t-z_{\ell}(t)}\right)^{\frac{1}{\sqrt{3}}}\frac{1}{\sqrt{1-\frac{t^2}{z_{\ell}(t)^2}}}=\frac{t}{t_0}.
\end{equation}
We find that this expression matches exactly the position at which our numerical solution keeps behaving unaffected by the blackbody turn-off. Notice that at late times, this expression admits the asymptotic behavior
\begin{equation}
    \frac{z_\mathrm{\ell}(t)}{t}=1-\left(\frac{t_0}{t}\right)^{\frac{2\sqrt{3}}{2+\sqrt{3}}}2^{\frac{2-\sqrt{3}}{2+\sqrt{3}}}.
\end{equation}
This shows that the sound horizon never catches up with a light signal emitted simultaneously, since the delay with the light signal keeps growing with time.

On the other hand, a fluid element that had been emitted, say, at the initial time $t_i$, moves at every instant with velocity $v(z,t)$, so it is initially slower and at some point will be caught in the tail of the motion which feels the lack of pressure behind the expanding fluid and is slowed down. We can prove this picture by solving the equation $dz/dt=v(z,t)$ with the initial condition $z(t_i)=0$. The calculation follows the exact same steps as above, so we only report the asymptotic result for the position of the fluid element at very late times
\begin{equation}
    \frac{z(t)}{t}=1-\left(\frac{t_i}{t}\right)^{\frac{2}{\sqrt{3}+1}}2^{\frac{\sqrt{3}-1}{\sqrt{3}+1}}.
\end{equation}
From here we see that, independently of how much earlier the fluid element has been emitted, that is of how much smaller is $t_i$ compared to $t_0$, in the end all fluid elements will be caught up behind the sound horizon. 

Thus, as time goes by, an ever larger part of the fluid will fall behind the sound horizon: an alternative way to show this is to integrate the lab-frame energy density $(4\gamma^2-1)\rho/3$, using Eqs.~\eqref{eq:self_similar_velocity} and~\eqref{eq:self_similar_density}, between $z_\mathrm{\ell}(t)$ and $z=t$, to determine the total amount of energy that remains unperturbed. The integral can be performed analytically and, in the limit of $t\to\infty$, gives
\begin{equation}
    \int_{z_\mathrm{\ell}(t)}^{t}\frac{4\gamma^2-1}{3}\rho\,dz\simeq0.21e_\mathrm{eq}\frac{3+2\sqrt{3}}{3}t\left(\frac{t_0}{2t}\right)^{\frac{4}{2+\sqrt{3}}}.
\end{equation}
Thus the amount of energy of the fluid that remains unaffected by the blackbody turning off decreases as $t^{\frac{\sqrt{3}-2}{\sqrt{3}+2}}$. This result confirms that in plane geometry, ultimately all of the emitted fluid will be caught in the tail, because it has not accelerated fast enough to escape the sound horizon $z_\mathrm{\ell}(t)$; once it crosses this surface, it can finally feel the effect of the blackbody turning off.

While the fluid in the tail behind the emission cannot be described analytically in the setup described above, we can gain some intuition about its behavior by taking the extreme example of a semi-infinite column of fluid with comoving energy density $\rho_0$, for $z>\zs$, moving with the speed of sound -- as is the case for the self-similar solution when $t\to \infty$ -- which is suddenly released free at $z=0$. The lack of pressure support causes the fluid to expand to negative $z$, even though it was initially moving to the right with the speed of sound. The information about the lack of pressure support can only propagate to the right of the sound horizon given by $dz_\mathrm{\ell}/dt=(v+\vs)/(1+v \vs)$, where in this case $v=\vs$ in the unperturbed motion, so this equation trivially integrates to $z_\ell=\sqrt{3}t/2$. Thus, for $z>\sqrt{3}t/2$ the fluid continues its uniform sonic motion.

Behind the sonic horizon, a rarefaction wave launches to the right, caused by the depression behind the fluid. The absence of characteristic length scales implies that a self-similar solution can be found, with the condition that $v=\vs$ at $z_\ell=\sqrt{3}t/2$. One can easily verify that the velocity profile
\begin{equation}
    v=\frac{\sqrt{3}z-t}{z-\sqrt{3}t}
\end{equation}
satisfies the equations of motion and the boundary condition. This velocity profile is valid, of course, for $z<\sqrt{3}t/2$, and furthermore for $z>-t$, where $v=-1$, where the fall-back motion that we had anticipated above becomes luminal. By integrating the equations of motion, we can also find the density profile
\begin{equation}
    \rho(z,t)=\rho_0\left(\frac{t+z}{t-z}\right)^{2\sqrt{3}} \left(\frac{2-\sqrt{3}}{2+\sqrt{3}}\right)^{2\sqrt{3}},
\end{equation}
chosen so that $\rho(\sqrt{3}t/2,t)=\rho_0$. Thus the fluid originally in sonic motion is gradually entranced by the lack of pressure support into a tail with this self-similar profile, which smoothly goes to $0$ at $z=-t$.

\subsection{Spherical geometry}

In spherical geometry, the fluid accelerates immediately nearly to the speed of light. Using the same time dependence for the blackbody temperature, we show in Fig.~\ref{fig:snapshots_td} (right panels) the evolution of the flow. It is still true that a part of the fluid proceeds completely unaware that the blackbody behind it has turned off and that a tail develops behind it. However, as we will now prove, there is a crucial difference from the plane case: here, this tail does not feed on the fluid in front of it that is performing its unperturbed radial motion, and therefore the bulk of the fluid proceeds maintaining its shape and thickness, without leading to a sizable increase in the width of the neutrino shell and therefore in the duration of the fireball. The reason for this difference is that in spherical geometry, the fluid moves much closer to the speed of light much closer to the emitting surface compared to the plane-parallel case.

To prove our claim, we can deduce again the characteristic line to the right of which the fluid remains unperturbed, using the definition
\begin{equation}
    \frac{dr}{dt}=\frac{v(r)+v_s}{1+v_s v(r)},
\end{equation}
with $r(t=t_0)=\rs$ and $v(r)$ is now the steady velocity profile given implicitly by Eq.~\eqref{eq:velocityprofile}. Notice that, while in the plane case the velocity profile always depends on the ratio $z/t$, here it is steady and only depends on $r$. This equation can be integrated numerically to give $r_\mathrm{\ell}(t)$, the position to the right of which the fluid proceeds unaffected. We find that the solution to this equation nearly immediately evolves to the asymptotic profile
\begin{equation}
    r_\mathrm{\ell}(t)\simeq \rs+(t-t_0)-0.065\,\rs,
\end{equation}
so it lags behind a light signal emitted from the surface only by $0.065\,\rs$ -- the delay accumulated in the initial phase of acceleration from sonic to near-luminal motion. In fact, this delay that we find numerically can be analytically expressed in terms of a rather complex combination of hypergeometric functions and Gamma functions; we do not reproduce the complicated expressions which however exactly match with the numerical integration.

On the other hand, a fluid element emitted at a time $t_i$ and moving with velocity $v(r)$ in the unperturbed region has an asymptotic position at a much later time $t$ which we find numerically to be
\begin{equation}
    r_\mathrm{f}(t)\simeq \rs+(t-t_i)-0.25\,\rs.
\end{equation}

We now see the crucial difference between spherical and plane geometry. In plane geometry, fluid elements never catch up with the speed of light, and the lag behind a light signal keeps growing in time. All of the fluid is ultimately entranced by the sound horizon, and slowed down by the depression behind the fluid. In spherical geometry, the fluid elements accumulate delay only very early, while accelerating, but the lag saturates to a finite value. Therefore, fluid elements emitted sufficiently early before the blackbody turns off will never be caught by the sound horizon and continue to propagate unaffected. Specifically, any fluid element that has been emitted before the time $t_i=t_0-0.19\,\rs$ will never fall below the sound horizon. More generally, it follows that at any given instant fluid elements emitted more than $\Delta t=0.19\,\rs$ earlier cannot be affected by any perturbation at the emitting surface. Integrating numerically the equation $dr/dt=v(r)$, we find that such fluid elements are beyond the sonic horizon $r_\mathrm{h}\simeq 1.13\,\rs$. Therefore, if the emission surface is at $\rs=10$~km, the sound horizon is only 1.3~km further out.

Because in a real SN we have a hierarchy that the signal duration is orders of magnitude larger than $\rs$, it follows that only a negligible part of the emission can actually be caught in the part of the motion that is slowed down by the blackbody turning off. The fact that most of the fluid remains unaffected can also be deduced by computing, as we did for the plane case, the total energy contained in the region between the sound horizon $r_\mathrm{\ell}(t)$ and $r=\rs+t$, the position of the front; since asymptotically at large radii the lab-frame energy density decreases as $e\propto r^{-2}$, it follows that this energy is 
\begin{equation}
    \int_{r_\mathrm{\ell}(t)}^{t}e\,4\pi r^2 dr\propto t_0+0.065\,\rs,
\end{equation}
which is time-independent and therefore not subject to further losses of fluid drifting to the tail region.

\section{Discussion and Summary}
\label{sec:discussion}

Motivated by the possible existence of large $\nu$SI and the question of how they would impact SN physics, we have studied radiative energy transfer as well as outflow from a hot radiating body representative of a collapsed SN core. The long and short of our findings is that the fluid nature of a self-coupled relativistic neutrino gas has surprisingly little impact on these phenomena, although quantitatively, numerical modifications arise for the Rosseland mean opacities or the outflow rate from the neutrino sphere.

Our main approach was to use the fluid equations of energy and momentum conservation, obviating the need for a detailed understanding of its internal state. As long as the interaction rates do not depend on energy, it is irrelevant, for example, if the fluid reaches chemical equilibrium by number-changing $\nu$SI, although otherwise numerical differences arise for the transport coefficients because of different averaging procedures over energy.

Our main result was to match the diffusion regime deeply inside the radiating body with the free expansion outside, leading to a consistent treatment of the outflow into space.
Our steady-state solution resembles the steady wind solution of Ref.~\cite{Chang:2022aas}, but we have taken it one step further in that we have matched the hydrodynamical flux to a thermal source. In this way, the properties of the source feed through all the way to the detectable neutrino signal.

We have completely dismissed the picture of a sudden release of a blob of neutrino fluid that was initially used in Ref.~\cite{Dicus:1988jh} to illustrate that $\nu$SI do not prevent neutrinos from streaming away. It was also proposed in Ref.~\cite{Chang:2022aas} as an alternative to steady outflow, but without explaining how the required initial condition for the entire blob would be engineered by SN-related physics.

We have assumed throughout that neutrinos behave as a fluid in every stage of the emission, but this condition has different quantitative meaning in different regions. In the diffusion regime deep inside the PNS, neutrinos behave as a fluid if they equilibrate between collisions with nuclei, implying the strong assumption $\lambda_{\nu\nu}\ll \lambda_{\nu N}$, with detailed modifications by neutrino degeneracy.

As for the outflow from the neutrino sphere, where by definition $\lambda_{\nu N}\simeq \rs$, the fluid nature of neutrinos here requires the much less stringent condition $\lambda_{\nu\nu}\ll \rs$. This is, in order of magnitude, the condition that is sometimes invoked to define the ``trapping'' region, for example in Ref.~\cite{Fiorillo:2022cdq}. We may then expect that couplings larger than the trapping region by, say, one order of magnitude would be described by the fluid dynamics discussed here.

For fireball propagation, the requirement is yet weaker: the MFP $\lambda_{\nu\nu}$ must be much shorter than the thickness of the propagating shell. On the other hand, as the fireball expands, the rest-frame number density decreases, so $\lambda_{\nu\nu}$ increases and of course at some point it will become larger than the thickness of the shell. The details of this decoupling are discussed in our companion paper~\cite{Fiorillo:2023ytr}. (See also Refs.~\cite{Diamond:2023scc,Diamond:2023cto} for a similar decoupling in the context of axion-sourced photon fireballs.)

The exact meaning of fluid behavior vs.\ traditional one would be important for a detailed study for 
particle-physics parameters where neutrinos might behave like usual radiation deeply inside the PNS, but as a fluid for outflow and initial fireball propagation. Our finding that fluid behavior causes only small modifications in any phase remains unchanged if, for given parameters, the fluid behavior only occurs in the later phases.

On the technical level, our results agree with the foundational ones of Dicus et al.\ \cite{Dicus:1988jh} concerning the sudden expansion of a fluid in plane-parallel geometry, although the self-similar nature of the solution was somewhat obscured in their expressions. However, we significantly advance on those results by considering the diffusive dynamics of the interior of the PNS, as well as extending results to spherical geometry. Besides, even in plane geometry, these authors consider an infinite reservoir of fluid, and therefore never discuss what happens to the tail of the signal after the reservoir is exhausted.

For spherical geometry, the dynamics outside the PNS quickly stabilizes into a steady solution, whose velocity profile coincides with the ``steady wind'' discussed in Ref.~\cite{Chang:2022aas}. However, while Ref.~\cite{Chang:2022aas} proposed that this solution may require unique conditions within and outside the PNS to actually be realized, we show that an initial condition corresponding to the PNS beginning to radiate in a vacuum does actually relax quickly to the steady state. This in turn descends from the sonic nature of the outflow, by which the solution close to the PNS cannot depend sensitively on the initial, transient emission far from it. In addition, compared to Ref.~\cite{Chang:2022aas}, we are able to match this solution with the 
diffusive thermal source and explicitly predict the energy flux emitted by the PNS, albeit for a simplified source model in the form of an isothermal body.

In the diffusion limit, Cerde{\~n}o et al.\ \cite{Cerdeno:2023kqo} have proposed that the energy flow would be identical between a fluid and standard radiation, using an argument of momentum conservation in binary neutrino collisions. Such an argument is in itself not convincing, since the energy flow depends on the collisions \textit{with the nucleons}. In turn such collisions depend on the angular and energy distribution of neutrinos, which can be changed by secret interactions even if momentum and energy are conserved in individual neutrino-neutrino collisions. We do find that the energy transport is identical, albeit only for energy-independent $\lambda_{\nu N}$, but we clarify that a kinetic approach to gain insight into the internal fluid interactions can be too simplistic, and the reasons for the similar transport go beyond momentum conservation. Still, we broadly agree with their conclusion for the dynamics inside the PNS.

We have also discussed the ``tail'' of the neutrino signal, emitted when the blackbody emission starts to decrease. We address the somewhat counterintuitive idea of a shell of neutrino fluid that propagates without expanding, in spite of the lack of pressure support behind it. In plane geometry, indeed the whole of the fluid initially propagating as a shell is ultimately slowed down by the smaller density behind it, resulting in a stretch of the thickness of that shell. However, in spherical geometry, this phenomenon is much less pronounced, and ultimately irrelevant for observable features. The reason is that the information about the blackbody turn off cannot catch up with the already emitted fluid, which has accelerated to essentially the speed of light. Since the fluid speed becomes nearly luminal at distances of the order of $\rs$, the radius of the surface, while the duration of the signal is orders of magnitude larger, only a negligible part of the fluid is actually affected by the blackbody turning down.
This phenomenon is familiar from the usual fireball expansion which thus fundamentally depends on spherical geometry as opposed to a plane-parallel model.

Thus, while the boundary condition for the fluid outflow at the neutrino sphere is the same for plane and spherical geometry, the subsequent propagation differs strongly. In spherical geometry, once the fluid becomes essentially luminal, the fluid nature of the neutrinos becomes essentially irrelevant from the point of view of their angular distribution. In this sense, one would not expect a strong modification of the SN burst duration observable at Earth: The burst duration is essentially set by the outflow speed at the PNS and not by the fireball propagation. On the other hand, the internal fluid dynamics (chemical equilibrium or not) may in principle have some impact on the observable signal properties. 

We will return to these more phenomenological questions in a companion paper \cite{Fiorillo:2023ytr}, where we use the technical input from the present more abstract study.

\section*{Acknowledgments}
We thank Shashank Shalgar, Irene Tamborra, Mauricio Bustamante, Po-Wen Chang, Ivan Esteban, John Beacom, Todd Thompson, Christopher Hirata, and Thomas Janka for informative discussions and/or comments on the manuscript. DFGF is supported by the Villum Fonden under Project No.\ 29388 and the European Union's Horizon 2020 Research and Innovation Program under the Marie Sk{\l}odowska-Curie Grant Agreement No.\ 847523 ``INTERACTIONS.'' GGR acknowledges partial support by the German Research Foundation (DFG) through the Collaborative Research Centre ``Neutrinos and Dark Matter in Astro- and Particle Physics (NDM),'' Grant SFB-1258-283604770, and under Germany’s Excellence Strategy through the Cluster of Excellence ORIGINS EXC-2094-390783311. EV acknowledges support by the European Research Council (ERC) under the European Union’s Horizon Europe Research and Innovation Program (Grant No.\ 101040019). This article  is based upon work from COST Action COSMIC WISPers CA21106, supported by COST (European Cooperation in Science and Technology).

\appendix

\section{Energy-dependent interaction}\label{app:energy_dependent}

In the main text, we have usually assumed that the neutrino interaction with the background medium does not
depend  on energy (``gray atmosphere model'') and that only emission and absorption contribute, not scattering. We here briefly explore some consequences of the inevitable energy dependence, for neutrinos roughly quadratic.

In the diffusion regime of Sec.~\ref{sec:Diffusion}, we have found the usual Rosseland mean in Eq.~\eqref{eq:Rosseland} and fluid equivalent in Eq.~\eqref{eq:Rosseland-Fluid}. If we assume the MFP varies as
\begin{equation}\label{eq:quadratic}
    \lambda_{\nu N}=\lambda_0\,\left(\frac{\epsilon_0}{\epsilon}\right)^2,
\end{equation}
these averages are
\begin{subequations}
\begin{eqnarray}
     \bar\lambda_{\rm rad}&=&~~\frac{5}{7\pi^2}\,\frac{\lambda_0\epsilon_0^2}{T^2}~\,
     \simeq\lambda_0\left(\frac{\epsilon_0}{3.72\,T}\right)^2,
     \\
      \bar\lambda_{\rm fluid}&=&\frac{147}{310\,\pi^2}\,\frac{\lambda_0\epsilon_0^2}{T^2}
      \simeq\lambda_0\left(\frac{\epsilon_0}{4.56\,T}\right)^2,
\end{eqnarray}    
\end{subequations}
implying $\bar\lambda_{\rm fluid}/\bar\lambda_{\rm rad}=1029/1550\simeq0.664$ and thus a somewhat reduced MFP for the fluid.

In the diffusion limit, the fluid bulk velocity $v\ll1$ is nonrelativistic, allowing one to use the same interaction rate for energy and momentum exchange, whereas in the decoupling region near the PNS surface, $v$ reaches the velocity of sound and one has to be more systematic. The energy extracted from the nucleon fluid per unit time and volume by collisions (emission and absorption) is
\begin{equation}
    \left(\frac{\partial e}{\partial t}\right)_{\rm coll}=-6\int \frac{d^3\mathbf{p}}{(2\pi)^3}
    \epsilon_\mathbf{p}\,\frac{f-f^{\rm th}}{\lambda_{\nu N}},
\end{equation}
where $f$ is the neutrino occupation number, $f^{\rm th}$ its thermal counterpart appropriate to the surrounding medium $T$, $\epsilon_\mathbf{p}$ is the neutrino energy, and the factor 6 accounts for the number of neutrino species. Similarly, the gain in the neutrino number is
\begin{equation}
    \left(\frac{\partial n}{\partial t}\right)_{\rm coll}=-6\int \frac{d^3\mathbf{p}}{(2\pi)^3}
    \,\frac{f-f^{\rm th}}{\lambda_{\nu N}}.
\end{equation}
Finally, the gain in momentum along the axis $z$ is
\begin{equation}
    \left(\frac{\partial (e\xi)}{\partial t}\right)_{\rm coll}
    =-6 \int \frac{d^3\mathbf{p}}{(2\pi)^3} \epsilon_\mathbf{p}\cos\theta\,
    \frac{f-f^{\rm th}}{\lambda_{\nu N}}.
\end{equation}
where $\cos\theta$ is the angle of the neutrino with the $z$ axis. If the MFP does not depend on energy, these expressions lead to the exchange terms adopted in the main text.

If instead the relaxation rate does depend on energy, these exchange terms would become more complex, non-linear functionals of the energy density, momentum density, and neutrino number density. As an explicit example, we use quadratic energy dependence of Eq.~\eqref{eq:quadratic}. Moreover, we assume that number-changing reactions in the neutrino fluid are so fast that both kinetic and chemical equilibrium obtain and the fluid is fully described by its bulk velocity $v$ and its internal temperature $T_\nu$ in its rest frame, not to be confused with the surrounding matter temperature $T$ in the matter rest frame. The integration then leads
to the form for the energy and momentum exchange term
\begin{equation}
    \left(\frac{\partial e}{\partial t}\right)_{\rm coll}=-\frac{1}{\lambda_0\epsilon_0^2}\,
    \frac{31 \pi^4}{84}
    \left[\frac{5+10v^2+v^4}{5(1-v^2)}\,T_\nu^6-T^6 \right].
\end{equation}
A similar integration for the momentum-exchange term leads to
\begin{equation}
    \left(\frac{\partial \left(e\xi\right)}{\partial t}\right)_{\rm coll}=
    -\frac{1}{\lambda_0\epsilon_0^2}\,\frac{31 \pi^4}{84}\,\frac{2v(5+3v^2)}{5(1-v^2)}\,T_\nu^6.
\end{equation}
By inverting the standard thermodynamic relations, these can be expressed in terms of $e$ and $\xi$ as nonlinear expressions. 

The corresponding equations for the steady outflow are
\begin{equation}
    \partial_z F=\left(\frac{\partial e}{\partial t}\right)_{\mathrm{coll}}
\quad\hbox{and}\quad
\partial_z \mathcal{M}=\left(\frac{\partial (e\xi)}{\partial t}\right)_{\mathrm{coll}}.
\end{equation}
They can be solved numerically with the usual initial condition that the neutrino fluid is in thermal equilibrium deep in the body. Our result that the maximum outflow velocity from a hot body is the speed of sound is entirely independent of the specific form of the exchange term, since it only depends on the property of the fluid equations that the characteristics have the slope of the speed of sound. Surprisingly, the numerical integration reveals that, for this modified energy-momentum exchange term, also the energy density at sonic outflow from the body is still equal to $e\simeq 0.35e_\mathrm{eq}$. While we cannot draw from this the general conclusion that \textit{any} energy dependence of the MFP would lead to identical results, this is certainly suggestive that our assumption of energy-independent MFP does not strongly impact our conclusions.

\section{Self-consistent temperature profile}

\label{sec:Milne}

In the main text, we have always assumed the radiating body to have a prescribed temperature profile which, moreover, was taken to be isothermal. In a realistic astrophysical situation, on the other hand, this profile is the result of energy transfer by neutrinos and other agents, such as convection, as well as the hydrodynamic solution for the background medium. Usually the temperature profile decreases toward the surface and the emerging radiation is a superposition of blackbody spectra emitted from different depths. Typically it is a good approximation to think of the radiation as emerging from the optical depth $\tau=2/3$ with a flux corresponding to the Stefan-Boltzmann Law, evaluated with the temperature and radiating surface at that depth
\cite{Caputo:2022rca}. If the interaction rate depends strongly on energy, as is the case for neutrinos, the idea of an emitting ``neutrino sphere'' strongly depends on energy and thus becomes a rather approximate concept.

In normal radiative transfer in plane-parallel geometry, however, a self-consistent matter profile is not necessary because one can measure distance along $z$ in units of the MFP and depth is measured as the dimensionless optical depth $\tau$ of the radiation. This situation is equivalent to using a homogeneous matter profile that ends abruptly at the surface, even though geometrically it may follow a power-law profile, for example, that geometrically never ends. If there are other scales in the problem, this approach is no longer exact, for example in spherical geometry, where the radiation decoupling layer need not be thin compared to the radius. 

However, to develop an impression of the difference between standard radiation and fluid, we now consider a plane-parallel situation and search for a self-consistent temperature profile $T(\tau)$ or rather, of $e_\mathrm{eq}(\tau)\propto T^4(\tau)$, where $e_\mathrm{eq}$ as usual represents the neutrino energy density in LTE prescribed by $T(\tau)$, where $\tau$ is the optical depth.

The temperature profile is self-consistent in a stationary state if there is no energy exchange between matter and the neutrino fluid, implying that the lab-frame neutrino energy density $e(z)=e_\mathrm{eq}(z)$. In turn, this means that the energy flux $F=e\xi$ is constant: $\partial_z(e\xi)=0$. This formulation of the problem was also used in the wind outflow scenario of Ref.~\cite{Chang:2022aas}, providing the matching with the dynamics inside the supernova.

\begin{figure}
    \centering
    \includegraphics[width=\columnwidth]{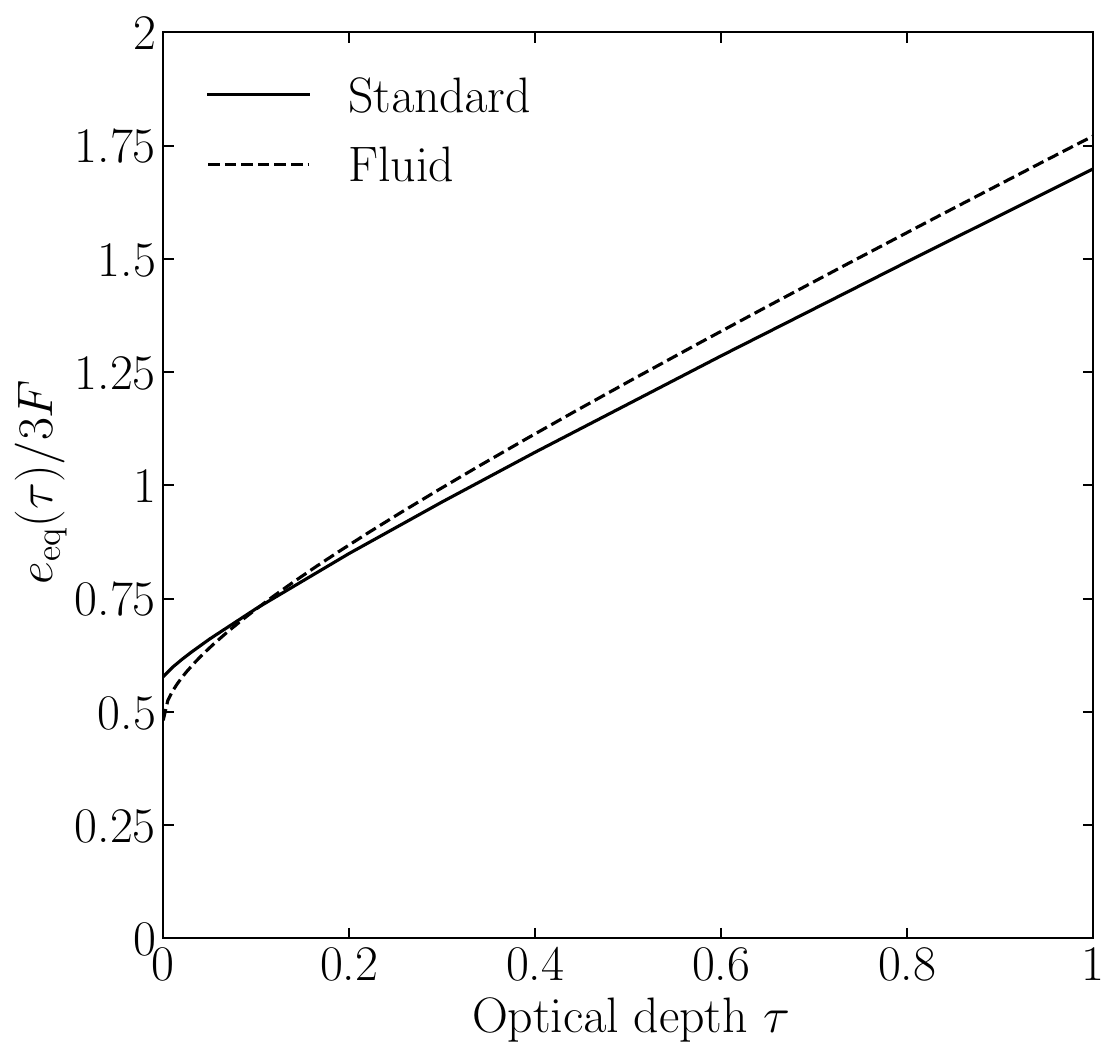}
    \caption{Self-consistent temperature profile as a function of optical depth, for the standard noninteracting case (solid) and the interacting  fluid case (dashed). $F$ is the chosen value for the constant energy flux in this problem.}
    \label{fig:self_consistent_profile}

\end{figure}

The equation of momentum conservation Eq.~\eqref{eq:momentum-conservation} for the neutrino fluid then gives
\begin{equation}\label{eq:momentum-conservation-sc}
    \partial_z\left[\frac{5-2\sqrt{4-3\xi^2}}{3\xi}\right]=-\frac{1}{\lambda_{\nu N}}.
\end{equation}
Let us call $\tau=-\int_0^z \frac{dz'}{\lambda_{\nu N}(z')}$ the optical depth measured from the surface inside the material. Then Eq.~\eqref{eq:momentum-conservation-sc} is easily integrated and yields
\begin{equation}
    5-2\sqrt{4-3\xi^2}=3\xi(\alpha+\tau).
\end{equation}
Here $\alpha$ is an integration constant, to be chosen according to a boundary condition which we may take to be the fluid bulk velocity at the emission point. We may take this condition to be a sonic outflow, i.e., $\xi=\xs=2\sqrt{3}/5$ at $\tau=0$, which gives \smash{$\alpha=\sqrt{3}/2$}. Then the velocity profile becomes
\begin{equation}
    \xi(\tau)=\frac{6}{5\sqrt{3}+10\tau+8\sqrt{\tau\left(\sqrt{3}+\tau\right)}}.
\end{equation}
In turn, the energy density behaves as
\begin{equation}
    e_\mathrm{eq}(\tau)=e(\tau)=\frac{F}{\xi(\tau)},
\end{equation}
shown in Fig.~\ref{fig:self_consistent_profile} as a dashed line. Here the constant energy flux $F$ is a chosen parameter of our problem. Deep inside the material ($\tau\to\infty$) this is $e_\mathrm{eq}(\tau)=3F\tau$, providing a linearly increasing profile with depth. Close to the emission point, the profile drops with a vertical asymptote as the speed approaches the speed of sound, corresponding to the result obtained also in the main text that sonic emission is the maximum possible outflow velocity.

We may compare this result with the corresponding case for a gray atmosphere with noninteracting radiation as the dominant heat transfer mechanism. This amounts to determining the solution of the problem
\begin{equation}
    \mu \partial_\tau f(\mu,\tau)=-f^{\rm th}(\tau)+f(\mu,\tau),
\end{equation}
where $f(\mu,\tau)$ is the distribution function of neutrinos in the lab frame with $\mu=\cos\theta_z$ the cosine of the angle with the $z$ axis and $\tau$ is the optical depth. Here $f^{\rm th}(\tau)$ is the equilibrium distribution function that neutrinos would assume in LTE. In turn, the local temperature of the medium should be self-consistently determined in such a way that the energy flux carried by neutrinos
\begin{equation}
    F=6\int f \mu \frac{d^3\mathbf{p}}{(2\pi)^3}
\end{equation}
is constant. Here we have included a factor 6 for six species, but this is not crucial in the following. The problem should be solved with the condition that at the surface ($\tau=0$) there are no incoming particles, i.e., $f(\mu,0)=0$ for $\mu<0$.

The solution to this problem can be found by analogy, after one notices that in a scattering atmosphere, where neutrinos are not emitted but only scattered, the energy flux is an automatic invariant of motion. Therefore, one needs only solve the corresponding problem for a scattering atmosphere
\begin{equation}
    \mu \partial_\tau f(\mu,\tau)=-\frac{1}{2}\int_{-1}^{+1}d\mu' f(\mu',\tau)+f(\mu,\tau)
\end{equation}
and finally identify the source function 
\begin{equation}
    f^{\rm th}(\tau)=\frac{1}{2}\int_{-1}^{+1}d\mu f(\mu,\tau).
\end{equation}
This is the well-known Milne problem, whose solution can be found to a very good approximation using Gaussian sums~\cite{chandrasekhar2013radiative}. One usually finds the solution in a conventional form involving a tabulated Hopf function $q(\tau)$; in our notation, this form is
\begin{equation}
    e_\mathrm{eq}(\tau)=6\int f^{\rm th}(\tau)\frac{d^3 \mathbf{p}}{(2\pi)^3}=3F\bigl[\tau+q(\tau)\bigr].
\end{equation}
The function $q(\tau)\simeq2/3$ is nearly constant over the entire range. $e_\mathrm{eq}(\tau)$ based on this constant value is called the Milne-Eddington approximation and can be found by simple heuristic arguments, e.g., in Ref.~\cite{Esposito:2010zz}.

We show the standard $e_\mathrm{eq}(\tau)$ in Fig.~\ref{fig:self_consistent_profile} as a solid line, where we use the Hopf function $q(\tau)$ as tabulated in
Ref.~\cite{1960ApJ...132..509K}. A representation in terms of integral expressions is provided in Eq.~(3-79) of Mihalas~\cite{Mihalas:1978}. In the limit $\tau\to\infty$, one finds~\cite{Mihalas:1978}
\begin{equation}
    q(\infty)=\frac{6}{\pi^2}+\frac{1}{\pi}\int_{0}^{\pi/2}d\theta
    \left(\frac{3}{\theta^2}-\frac{1}{1-\theta\cot\theta}\right)
    \simeq 0.710.
\end{equation}
Therefore, deep in radiating body, one finds $e_\mathrm{eq}(\tau)=3F(\tau+0.710)$, whereas in the fluid case, our corresponding result is
$e_\mathrm{eq}(\tau)=3F(\tau+0.866)$. As we had already found in Sec.~\ref{sec:Diffusion}, in the diffusion limit and for a gray-atmosphere model, the radiation and fluid cases provide identical energy transfer. Here this finding turns around in that they provide the same self-consistent linear profile for $e_\mathrm{eq}(\tau)$. On the other hand, this similarity breaks down close to the surface, where in the fluid case the flow becomes sonic and the energy density has a downward cusp.

One corollary of this comparison is that the energy flux from a nonisothermal body is given roughly by the usual Stefan-Boltzmann Law applied at optical depth $\tau=2/3$. The similarity between energy transfer by standard radiation or a fluid also manifests itself in this context. The findings of the main text are not limited to isothermal radiating bodies.

\bibliographystyle{bibi}
\bibliography{References}

\end{document}